\documentclass[letterpaper,english,aps, nofootinbib, prd, twocolumn, lengthcheck, superscriptaddress]{revtex4-2}
\usepackage[T1]{fontenc}

\usepackage[colorlinks=true,linktocpage=true,linkcolor=blue,citecolor=blue]{hyperref}
\usepackage{verbatim}
\usepackage{bm}
\usepackage{amsmath}
\usepackage{amssymb}
\usepackage{graphicx}
\usepackage{slashed}
\usepackage{wrapfig}
\usepackage{bbm}
\usepackage[caption=false]{subfig}
\usepackage{verbatim}
\newcommand \beq{\begin{eqnarray}}
\newcommand \eeq{\end{eqnarray}}

\newcommand{\Tr}{{\rm Tr}}

\newcommand{\Nc}{N_{\rm c}}
\newcommand{\Nf}{N_{\rm f}}

\newcommand{\lqcd}{\Lambda_{\rm QCD}}
\newcommand{\vp}{ {\bm p}}

\newcommand{\vP}{{\bm P}}

\newcommand{\la}{\langle}
\newcommand{\ra}{\rangle}

\newcommand{\calH}{\mathcal{H}}
\newcommand{\calS}{\mathcal{S}}  

\newcommand{\calA}{\mathcal{A}}
\newcommand{\calB}{\mathcal{B}}

\newcommand{\calI}{\mathcal{I}}

\newcommand{\calN}{\mathcal{N}}

\newcommand{\calV}{\mathcal{V}}
\newcommand{\calP}{\mathcal{P}}

\newcommand{\rmd}{\mathrm{d}}

\newcommand{\rme}{\mathrm{e}}

\newcommand{\up}{\uparrow}
\newcommand{\down}{\downarrow}

\newcommand{\tp}{ \tilde{p} }

\newcommand{\tP}{ \tilde{P} }

\newcommand{\blueflag}[1]{{\color{blue} #1}}

\begin{document}

\title{Stiffening of matter in quark-hadron continuity}


\author{Toru Kojo}
\affiliation{Key Laboratory of Quark and Lepton Physics (MOE) and Institute of Particle Physics, Central China Normal University, Wuhan 430079, China}  

\date{\today}

\begin{abstract}
We discuss stiffening of matter in quark-hadron continuity.
We introduce a model that relates quark wave functions in a baryon and the occupation probability of states for baryons and quarks in dense matter.
In a dilute regime, the confined quarks contribute to the energy density through the masses of baryons, but do not directly contribute to the pressure; hence, the equations of state are very soft.
This dilute regime continues until the low momentum states for quarks get saturated; this may happen even before baryons fully overlap, possibly at density slightly above the nuclear saturation density.
After the saturation the pressure grows rapidly while changes in energy density are modest, producing a peak in the speed of sound.
If we use baryonic descriptions for quark distributions near the Fermi surface, we reach a description similar to the quarkyonic matter model of McLerran and Reddy.
With a simple adjustment of quark interactions to get the nucleon mass, our model becomes consistent with the constraints from 1.4-solar mass neutron stars, 
but the high density part is too soft to account for two-solar mass neutron stars. 
We delineate the relation between the saturation effects and short range interactions of quarks, suggesting interactions that leave low density equations of state unchanged but stiffen the high density part.

\end{abstract}

\maketitle

\section{Introduction}
\label{sec:introduction}

How highly compressed baryonic matter transforms into quark matter has been a long standing question in quantum chromodynamics (QCD) \cite{Itoh:1970uw,Collins:1974ky}.
Considering the size of a baryon of $0.5-0.8$ fm, we expect the transition to take place at $n_B = 2-10n_0$ ($n_0 \simeq 0.16\,{\rm fm}^{-3}$: nuclear saturation density).
The difficulties to describe the transition lie in treatments of nonperturbative effects such as confinement, chiral restoration, or other strong correlation effects \cite{Fukushima:2010bq,Buballa:2014tba,Alford:2007xm}.
The lattice Monte Carlo simulations suitable for strong coupling regimes are not usable at finite density, 
while perturbative calculations based on the weak coupling picture are not applicable at $n_B \lesssim 40n_0$ \cite{Freedman:1977gz,Freedman:1976ub,Annala:2019puf,Gorda:2021kme,Gorda:2021znl,Fraga:2015xha,Kurkela:2014vha,Kurkela:2009gj,Fujimoto:2020tjc}.
The framework based on low energy nuclear physics \cite{Togashi:2017mjp,Akmal:1998cf,Drischler:2017wtt,Carlsson:2015vda,Drischler:2015eba,Lynn:2015jua,Carlson:2014vla,Oertel:2016bki} is reliable only to $n_B = 1.5-2n_0$ 
beyond which we must perform some extrapolation toward high density.

In spite of all these difficulties, the combined use of the above information and recent neutron star (NS) observations 
\cite{Arzoumanian:2017puf,Fonseca:2016tux,Demorest:2010bx,Cromartie:2019kug,Fonseca:2021wxt,Miller:2019cac,Riley:2019yda}  
allows us to get insights on the properties of dense matter (see, e.g., \cite{Kojo:2020krb} for a short review).
Recent analyses by NICER, including the radius measurements of $2.08M_\odot$ ($M_\odot$: solar mass) and $1.4M_\odot$ NSs, constraints from the NS merger event GW170817,  
and nuclear physics constraints, yielded the estimates of the radii $R_{2.08} \simeq R_{1.4} \simeq 12.4$ km \cite{Miller:2021qha,Riley:2021pdl}.
This small variation in the radii from $1.4$ to $2.08M_\odot$ NS suggests that the equation of state (EoS) for $n_B =2-5n_0$ should not contain substantial softening, but rather should get stiffened.
This feature disfavors strong first order phase transitions in the domain $n_B =2-5n_0$, although the weaker one is still possible.

Since the first discovery of $2M_\odot$ NS \cite{Demorest:2010bx}, a number of works have been devoted to the crossover description for hadron-to-quark phase transitions \cite{Masuda:2012kf,Masuda:2012ed,Masuda:2015kha,Kojo:2014rca,Baym:2017whm,Baym:2019iky,Ma:2019ery,Ayriyan:2021prr}.
Early works \cite{Masuda:2012kf,Masuda:2012ed,Masuda:2015kha} phenomenologically interpolated hadronic EoS at $n_B\lesssim 2n_0$ and quark matter EoS at $n_B \gtrsim 4n_0$; 
the resulting EoS is consistent with the existing NS constraints, and has a novel peak structure in the speed of sound $c_s = ( \partial \calP/\partial \varepsilon)^{1/2}$ where $\calP$ and $\varepsilon$ are pressure and energy density, respectively.
Later, such a peak structure was discussed as generic, by noting the contrast between stiffness of low and high density EoS \cite{Bedaque:2014sqa}; nuclear physics calculations suggest soft low density EoS which must get stiffened rapidly to pass the $2M_\odot$ mass constraints \cite{Tews:2018kmu,Drischler:2020fvz}.
This peak should have a mechanism specific to dense matter \cite{Pisarski:2021aoz,Hippert:2021gfs};  
in the finite temperature crossover from a hadron resonance gas to a quark-gluon plasma, the speed of sound has a dip, instead of a peak, as one can see from the lattice simulations \cite{Bazavov:2018mes}.

The microscopic mechanism for the emergence of the peak has been discussed by McLerran and Reddy (MR) \cite{McLerran:2018hbz}, who used the concepts of quarkyonic matter \cite{McLerran:2007qj}.
The quarkyonic matter is a quark matter with a baryonic Fermi surface, and the excitations are confined \cite{Glozman:2007tv,Hidaka:2008yy,McLerran:2008ua,Andronic:2009gj,Kojo:2009ha,Kojo:2010fe,Kojo:2011fh,Kojo:2011cn,Ferrer:2012zq,Tsvelik:2021ccp}.
In the MR model, they used a hybrid description in momentum space.
A matter at low density is dominated by baryons, but as density increases, the quark Fermi sea emerges at low momenta, pushing up the baryonic states to high momenta.
For suitable choices of parameters, baryons become relativistic at $n_B=1.5-3n_0$ with a substantial peak in $c_s$.
The advantage of the MR model is that relativistic baryons emerge by the quark Pauli blocking mechanism which is independent of details in nuclear forces, and hence the mechanism is qualitatively robust.
Several successful descriptions of the NS have originated from this framework \cite{Jeong:2019lhv,Duarte:2020xsp,Duarte:2020kvi,Han:2019bub,Zhao:2020dvu,Cao:2020byn,Margueron:2021dtx,Somasundaram:2021ljr}.

In this paper we discuss the stiffening of dense matter associated with the {\it saturation} of quark states at low momenta that may be regarded as the onset of the quark Fermi sea.
The preliminary discussion was given in Ref.\cite{Kojo:2019raj}, and this paper is the fuller version.
This work is basically a follow-up work of Ref.\cite{McLerran:2018hbz}, but contains new attempts and insights which can be potentially important.

First, we describe the crossover behavior using quark degrees of freedom only, starting with the description of a single baryon, proceeding to a baryonic matter, and then to a quark matter formation.
Although the descriptions are rather crude, this approach has the advantage over conventional hybrid descriptions 
where one uses quarks in one place and baryons (hadrons) in the other place. 
This removes the worries about double counting as well as the confusions associated with switching in degrees of 
freedom.\footnote{The double counting introduces serious problems into the field theoretic computations for the zero point energy in matter \cite{Blaschke:2013zaa}. The UV divergences appear from both elementary and composite particles. Consistent treatments of both contributions are mandatory to cancel the divergences in medium by the vacuum subtraction of the energy \cite{Kojo:2018usw}.}
The onset density of quark matter formation is related to the size scale of a baryon, and we found that the saturation begins to occur slightly above the nuclear regime, $1-3n_0$, even before baryon cores overlap. 
The softness in baryonic matter and stiffness in quark matter are described in a unified manner.
We also check how our quark descriptions are related to the MR model.

Second, we attempt to describe in-medium interactions at the level of quark descriptions.
This is potentially important, considering the difficulties to predict the high density behaviors of two-, three-, and more-body baryonic forces within purely baryonic calculations.
It has been known that a simple constituent quark model with one-gluon exchanges accounts for the baryon spectroscopy remarkably well \cite{DeRujula:1975qlm,Isgur:1979be}.
Recent lattice calculations for baryon-baryon interactions even show that the simple constituent quark picture correctly describes the observed patterns of baryon-baryon interactions, including the hard core repulsion among nucleons as well as baryon interactions with strangeness \cite{Oka:1980ax,Oka:1982qa,Park:2019bsz}.
These successes in describing semishort range correlations, at momentum scale $\sim 0.2-1$ GeV, 
give us a hope to build a unified description for the properties of matter from the baryonic to quark matter regime.
In this spirit, relevant interactions at $\sim 5-10n_0$, at lower density than for the perturbative regime, have been examined \cite{Kojo:2014rca,Baym:2017whm,Baym:2019iky,Song:2019qoh,Leonhardt:2019fua}.
The present framework solely based on quarks is suitable for the unified treatments of interactions from low to high densities.

In Sec.\ref{sec:quark_in_baryon}, we begin with quarks in a baryon, and
in Sec.\ref{sec:quark_in_baryonic_matter} we discuss quarks in a baryonic matter.
In Sec.\ref{sec:quark_matter_formation}, quark matter formation and the associated stiffening are described.
In Sec.\ref{sec:occ_baryon}, we mention how our descriptions are related to the MR model.
In Sec.\ref{sec:quantum_numbers}, we discuss the spin-flavor quantum numbers of baryons and how they fill the quark spin-flavor states.
In Sec.\ref{sec:interactions}, we discuss interactions for matter from the dilute baryonic to dense quark regimes
and examine what kind of interactions is suitable to describe the NS phenomenology.
Section \ref{sec:summary} is devoted to a summary.
In the Appendix, we discuss the phase space density in baryonic descriptions for the quark matter domain.


\section{Quarks in a baryon}
\label{sec:quark_in_baryon}

We consider how quark states are occupied as baryon density increases.
We 
postulate\footnote{ 
For a constituent quark model with $\Nc=3$ and the harmonic oscillator potential \cite{DeRujula:1975qlm,Isgur:1979be}, 
we can derive $Q_{\rm in}$ within simple analytic calculations. 
First we compute three quark wave functions $\Psi(\vp_1,\vp_2,\vp_3; \vP_B)$ and the corresponding probability function $|\Psi(\vp_1,\vp_2,\vp_3; \vP_B)|^2$. 
Then we integrate out two of momentum variables to get the single particle distribution $Q_{\rm in} (\vp ; \vP_B)$ in Eq.(\ref{eq:Q_in_Gauss}).
}
a distribution of quarks with the momentum $\vp$ which belong to a baryon with the momentum $\vP_B$; the form is given by
($\Nc$: number of colors)
\beq
Q_{\rm in} (\vp, \vP_B) = \calN \rme^{ -  \frac{1}{\, \Lambda^2 \,} \big( \vp - \frac{\, \vP_B \,}{\, \Nc \,} \big)^2 }\,,
\label{eq:Q_in_Gauss}
\eeq
where $\Lambda$ is the scale of the order of the QCD nonperturbative scale, $\lqcd \simeq 0.2-0.3$ GeV, related to the radius of a baryon as $\Lambda \sim R_{\rm baryon}^{-1}$, 
and $\calN$ is the normalization constant
\beq
\calN = \frac{\, 8\pi^{3/2} \,}{\, \Lambda^3 \,} \,,
\eeq
with which $\int_{\vp} Q_{\rm in} = 1$ 
(definition: $\int_{\vp} \equiv \int \frac{\rmd^3 \vp}{(2\pi)^3}$). 
The momentum distribution becomes broader for a smaller baryon radius.

We emphasize that $Q_{\rm in}$ is for a quark with a particular color, flavor, and spin.
When we compute baryonic quantities we have to sum contributions from all colors.
The sum of $\Nc$ quark momenta should be $\vP_B$, and the distribution satisfies this condition at the level of the expectation value,
\beq
\la \vP_B \ra = \Nc \int_{\vp} \vp Q_{\rm in} (\vp, \vP_B) \,.
\eeq
More realistically the probability distribution of a quark must be related to those of the other $\Nc-1$ quarks, 
but in this paper we consider only the averaged description for quark momentum distributions.

As the quark momentum has the broad distribution, the magnitude of momentum is substantial,
\beq
\bigg\la \bigg(\vp -\frac{\vP_B}{\Nc} \bigg)^2 \bigg\ra \sim \Lambda^2 \,.
\eeq
independent of $\vP_B$, and hence quarks can be energetic compared to the baryon kinetic energy or the nuclear scale of $O(1-10)$ MeV.
In baryonic matter, this disparity will be reflected as the large energy density but small pressure.

Below we approximate the baryon energy as the sum of the energies from $\Nc$ quarks. The energy of a quark in a baryon with momentum $\vP_B$ is
\beq
&&\la E_q (\vp) \ra_{\vP_B}
= \bigg\la E_q \bigg(\vp + \frac{\vP_B}{\Nc} \bigg) \bigg\ra_{\vP_B=0}
\nonumber \\
&& \simeq
\la E_q (\vp) \ra_{\vP_B=0}
+ \frac{\, \delta_{ij} \,}{\, 6 \,}  \bigg\la  \frac{\, \partial^2 E_q \,}{\, \partial p_i \partial p_j \,} \bigg\ra_{\vP_B=0} \bigg( \frac{\vP_B}{\Nc} \bigg)^2 + \cdots
\eeq
where the term linear in $\vP_B$ vanishes. It is important to note that the correction from finite $\vP_B$ is suppressed by $1/\Nc^2 \sim 1/10$, 
and the quark single particle energy is hardly affected by the baryon momentum until $P_B$ becomes very large, $\sim \Nc \Lambda$.

The simplest version of our model includes only a potential localizing quarks, and the resulting baryon energy is simply
\beq
E_B \simeq \Nc \la E_q (\vp) \ra_{\vP_B} \,.
\eeq
For instance, in a nonrelativistic quark model
\beq
E^{\rm NRq}_B = \Nc \bigg( M_q + \frac{\, \la \vp^2 \ra |_{\vP_B=0} \,}{\, 2M_q \,} \bigg) + \frac{\, \vP_B^2 \,}{\, 2 \Nc M_q \,} + \cdots\,,
\eeq
and the kinetic energy of the baryon is suppressed by the large baryon mass, $M_B \equiv E_B (P_B=0) $.

It is important to note that $M_B$ in our model is considerably larger than $\Nc M_q$ by the kinetic energy $\sim \Lambda$ of each quark.
If we use usual constituent quark mass of $M_q\sim 0.3$ GeV for up- and down-quarks, there must be attraction of $\sim \Lambda$ to keep the picture of $M_B \sim \Nc M_q$.
In Sec.\ref{sec:interactions}, we consider such short range interactions to modify the average single particle energy and to get the right baryon mass.

\section{Quarks in baryonic matter}
\label{sec:quark_in_baryonic_matter}

\subsection{Occupation probability of quark states}
\label{sec:Occupation_probability_of_quark_states}

We first write the occupation probability of quark states, $ f_q (p;n_B)$, with given color, flavor, and spin as
\beq
 f_q (p;n_B)  
=  \int_{\vP_B}\!  \calB (P_B; n_B) Q_{\rm in} (\vp, \vP_B) \,,
\label{eq:fq}
\eeq
where we wrote the occupation probability of baryon states as $\calB (P_B; n_B)$ for given flavor and spin\footnote{
For the moment we treat quarks and baryons as if they have no flavor and spin species, or discuss $u\up$ state in a $\Delta^{++}_{s_z=3/2} = (u_R\up, u_G\up, u_B\up )$ baryon. 
More details about spin-flavor quantum numbers will be addressed in Sec.\ref{sec:quantum_numbers}. }.
The expression means that $f_q$ in dense matter is obtained by summing up the quark occupation probability from each baryonic state.
Being probabilities, $ 0\le f_q \le 1$ and $0\le \calB \le 1$ must be satisfied.
Note that we have assumed that the probability depends only on the size of momenta, $p=|\vp|$ and $P_B = |\vP_B|$.

Integrating the quark momentum $\vp$ with the normalization $\int_{\vp} Q_{\rm in} = 1$,
we find
\beq
n_B = n_q^{R, G, B}  = \int_{\vp } f_q (p;n_B) = \int_{\vP_B}\!  \calB (P_B; n_B) \,,
\eeq
where $n_q^{R, G, B}$ is the quark density for a given color.
(We implicitly assumed the color neutrality condition $n_q^{R} = n_q^{G} = n_q^{B} $.
The quark number density is $n_q = \Nc n_B$.)

In dilute regime, we can neglect interactions among baryons as they are widely separated in space. 
Then baryons fill the states from low momenta with the probability $1$, as in an ideal gas;
($\Nf$: number of flavors)
\beq
\calB (P_B; n_B) = \theta( P_F - P_B) \,,~~~~~ 
n_B = 2\Nf \frac{\, P_F^3 \,}{\, 6\pi^2 \,} \,,
\eeq
where $\theta(x)$ is the step function, $P_F$ the Fermi momentum of baryons, 
and $2\Nf$ the factor from spins and flavors.

\begin{figure}[tb]
\begin{center}	
\vspace{-0.5cm}
	\includegraphics[width=8.8cm]{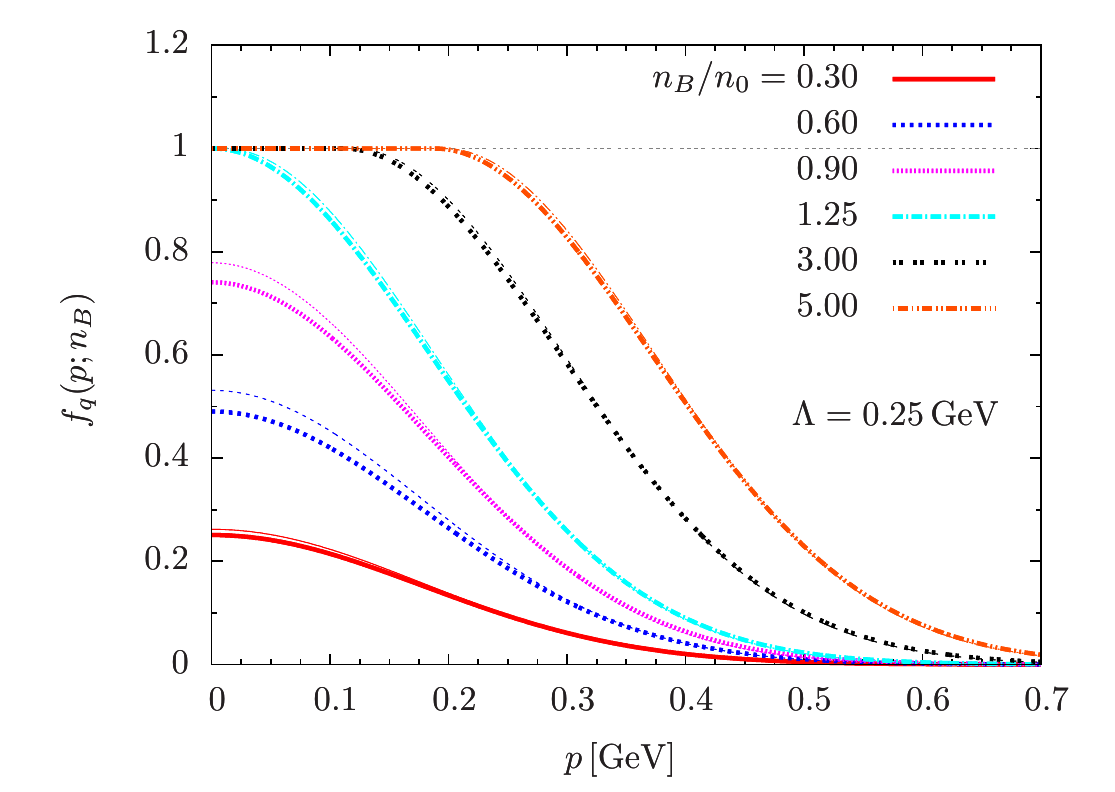}
\caption{ \footnotesize{The quark occupation probability $f_q(p;n_B)$, for various $n_B/n_0$. We took $\Lambda=0.25$ GeV and $\Nc=3$.
	The thin lines are the leading order of the $1/\Nc$ expansion. 
		} }
	\vspace{-0.5cm}
\label{fig:fq_for_pbf_lam}		
\end{center}
\end{figure}

Let us look at how $f_q$ evolves as $n_B$ increases. In the following, we rescale momenta,
\beq
\tp = p/\Lambda\,,~~~ \tP_B=P_B/\Nc \Lambda\,,~~~~~ \tP_F = P_F/\Nc \Lambda \,,
\eeq
with which $f_q$ can be written as
\beq
  f_q (p;n_B)  
&=& \calN \frac{\, (\Nc \Lambda)^3 \,}{\,  (2\pi)^2 \,} \int_0^{\, \tP_F \,} \tP_B^2 \, \rmd \tP_B ~  \rme^{ - \big( \tp^2 + \tP_B^2 \big) }
\nonumber \\
&&~~~~ \times \int^{1}_{-1}  \rmd \cos\theta ~ \rme^{ 2 \tp \tP_B \cos \theta } \,.
\label{eq:full_form_fq}
\eeq
We note that, in a dilute regime, $\tP_F\ll 1$ or $P_F \ll \Nc \Lambda$, so that the domain for the integral over $\tP_B$ is very small, canceling the overall factor $(\Nc\Lambda)^3$. 
As $\tP_B$ in the exponent of the integrand can be regarded as small, we expand them and find
\beq
f_q 
\simeq \calN \frac{\, P_F^3 \,}{\,  6 \pi^2 \,} \,  \rme^{ - \tp^2 }  \!
\bigg( 1+  \frac{\, -3+2 \tp^2 \,}{\, 5 \,}  \tP_F^2 + \cdots
\bigg) \,.	
\label{eq:fq_pF}
\eeq
Recalling $\calN \sim \Lambda^{-3}$, the overall size of $f_q$ is $\sim (P_F/\Lambda)^3$.
As for the shape, the leading order contribution maintains the Gaussian form same as in $Q_{\rm in}$.
The correction starts with the order of $1/\Nc^2 \simeq 1/10 $ for $\Nc=3$, so the large $\Nc$ should be a good approximation for the $\Nc=3$ case, 
except in situations where $p$ and $P_F$ are very large.
The numerical results of Eq.(\ref{eq:full_form_fq}) for various $n_B/n_0$ 
are shown in Fig.\ref{fig:fq_for_pbf_lam}, together with the leading contribution of the $1/\Nc$ expansion. We took $\Lambda= 0.25$ GeV and $\Nc=3$.
Here, we emphasize that we used the formula (\ref{eq:full_form_fq}) for only the curves up to  $n_B/n_0 = 1.25$;
for $n_B/n_0 > 1.25$, we used another expression Eq.(\ref{eq:fq_after_sat}) which is explained in the next section.
Here, $f_q$ for various $n_B/n_0$ are shown for $\Lambda= 0.25$ GeV and $\Nc=3$.

It is remarkable that the shape of $f_q$ hardly changes, while the size grows almost linearly in $n_B$.
Since $f_q$ is a probability, the growth in the size of $f_q$ must be terminated at $f_q=1$.
We call it {\it saturation of quark states}, or more simply, {\it quark saturation}.
The behavior of $f_q$ beyond $n_B/n_0 = 1.25$ is discussed in the next section.

In the large $\Nc$ limit, one can derive a number of simple expressions as we may neglect $\tP_F$ terms in Eq.(\ref{eq:fq_pF}). The $f_q$ takes the form
\beq
f_q (p;n_B)   \big|_{\Nc\rightarrow\infty} 
= \calN \frac{\, P_F^3 \,}{\,  6 \pi^2 \,} \,  \rme^{ - \tp^2 }  
= \frac{\, n_B \,}{\, n_B^c \,} \,  \rme^{ - \tp^2 }   \,,
\label{eq:f_q_largeNc}
\eeq
where the $p=0$ state gives the largest $f_q$.
The $p=0$ state gets saturated at $n_B^c = \Nf (P_F^c)^3/3\pi^2$; the corresponding baryon Fermi momentum is
\beq
1 = f(0; P_F^c) ~\leftrightarrow~ P_F^c \big|_{\Nc\rightarrow\infty} = \Lambda \bigg( \frac{\, 3 \sqrt{\pi} \,}{\, 4 \,}  \bigg)^{1/3} \simeq 1.1 \Lambda\,. \nonumber \\
\eeq

\begin{figure}[tb]
\begin{center}	
\vspace{-0.5cm}
	\includegraphics[width=8.8cm]{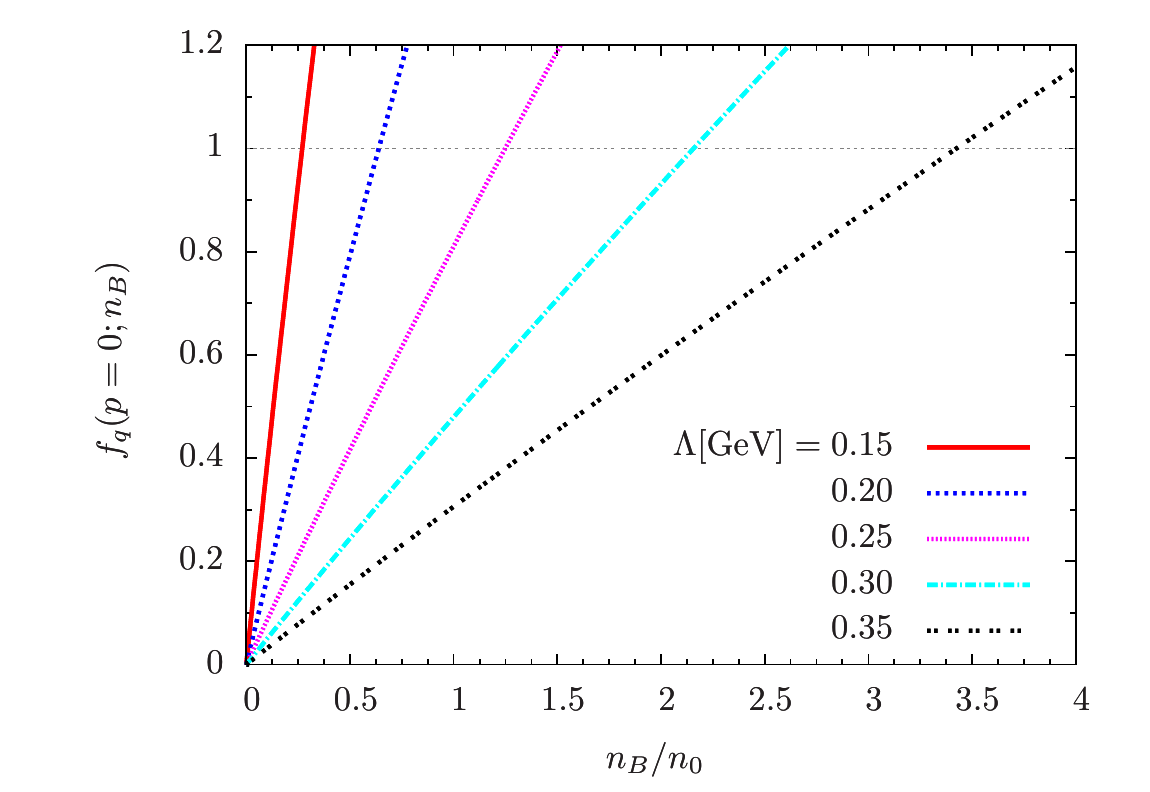}
\caption{ \footnotesize{The $n_B$ dependence of the quark occupation probability at zero momentum, $f_q(p=0;n_B)$, for $\Lambda=0.15, 0.20, 0.25, 0.30$, and $0.35$ GeV. We took $\Nc=3$.
The growth in $f_q(p=0;n_B)$ is almost linear in $n_B$.
		} }
	\vspace{-0.5cm}
\label{fig:FIG_fq_p0_lam}		
\end{center}
\end{figure}

In Fig.\ref{fig:FIG_fq_p0_lam} the $f_q (p=0; n_B) $ is shown as a function of $n_B$ for $\Nc=3$.
For a larger $\Lambda$, the baryon radius is smaller and the saturation of the $p=0$ state happens at larger $n_B$. 
It is important to note that, for $\Lambda=0.2-0.3$ MeV as the reasonable scale of baryon radii, the saturation takes place at $n_B = 0.5-2n_0$; 
this baryon density is within or close to the territory of conventional nuclear physics.
Such density range is well below the density $\sim 4-7n_0$ \cite{Baym:2017whm} where baryon cores begin to overlap.
The reason why this may happen would be that quark wave functions, in general, have broader extension than the average baryon size,
or the overlap of the meson cloud around a baryon take place before the cores overlap \cite{Fukushima:2020cmk}.

\subsection{Equations of state in dilute baryonic matter}
\label{sec:EOS_baryon}

Next we compute the EoS using the occupation probability discussed in the previous section. For simplicity we assume the quark energy of the form
\beq
E_q (p) = \sqrt{ p^2 + M_q^2} \,. 
\eeq
The energy density is computed as
\beq
\varepsilon (n_B) = 2\Nc \Nf \int_{\vp} E_q (p) f_q (p; n_B) \,.
\eeq
The chemical potential is computed as
\beq
\mu_B = \frac{\, \partial \varepsilon \,}{\, \partial n_B \,}
= 2\Nc \Nf \int_{\vp} E_q (p) 
\frac{\, \partial f_q (p; n_B) \,}{\, \partial n_B \,} \,.
\eeq
The expression is particularly simple in the large $\Nc$ limit; using Eq.(\ref{eq:f_q_largeNc}), we find
\beq
\varepsilon^{\Nc \rightarrow \infty} = \, n_B \Nc \calN \int_{\vp} E_q (p) \,  \rme^{-\tp^2} = n_B M_B^{\Nc \rightarrow \infty}  \,,
\eeq
so that $\mu_B = M_B$ at large $\Nc$. Accordingly the pressure is small, $\calP=\mu_B n_B - \varepsilon \sim n_B^{5/3}/M_B \sim 1/\Nc$, as expected from purely baryonic matter in an ideal gas regime.
This trend changes when quark states at low momenta get saturated, as we see in the next section.

\section{Quark matter formation}
\label{sec:quark_matter_formation}

We have seen that the quark states at low momenta begin to be saturated at $n_B^c \sim \Lambda^3$.
Beyond this critical density we can no longer use the ideal gas description of baryons.
In fact, the EoS behaves very differently before and after the saturation;
EoS are much stiffer after the saturation.

It is highly unconventional to discuss the baryon momentum distribution $\calB$ after the quark Fermi sea begins to be relevant.
In this situation, instead of starting with $\calB$, it is more intuitive to postulate the form of quark occupation probability $f_q$ for which we can implement the quark Pauli blocking easily.
In Sec.\ref{sec:occ_baryon}, we come back to the question of how $\calB$ looks like for the postulated form of $f_q$.

\subsection{A model of saturation}
\label{sec:model_saturation}

We assume that the occupation probability changes smoothly just after the saturation takes place. 
As a trial, we postulate the form
\beq
\hspace{-0.5cm}
f_q^{\rm after} 
= \theta \big( p_{\rm sat} - p \big)
+ \theta \big( p - p_{\rm sat} \big) \, f_q ( p - p_{\rm sat}; n_B^c) \,,
\label{eq:fq_after_sat}
\eeq
where $p_{\rm sat}$ is a function of $n_B$.
The first term is responsible for states occupied with the probability 1; the states to $p=p_{\rm sat}$ are saturated.
Meanwhile the second term is for states with $p>p_{\rm sat}$ which are only partially occupied.
Here we assume that the distribution $f_q$ at $n_B^c$ is shifted by the occupied level, 
so the states up to $\sim p_{\rm sat} +\Lambda$ can be occupied with substantial probability.
For $\Lambda \rightarrow 0$, the postulated form of $f_q$ is reduced to the form for the ideal quark gas. 
But we keep $\Lambda$ finite, assuming that quarks just after the saturation are not fully delocalized.
The behavior of $f_q$ after the saturation is shown in Fig.\ref{fig:fq_for_pbf_lam} for $n_B/n_0 > 1.25$ with $\Lambda = 0.25$ GeV.

While the postulated form of $f_q$ seems a small departure from $f_q$ in a baryonic matter, it has dramatic impacts on the EoS.
First, we note that the baryon densities  before and after the saturation are continuous, 
as we postulate the continuous changes in $f_q$.
The relation between the Fermi momentum and $p_{\rm sat}$ is given by
\beq
&&\frac{\, P_F^3 \,}{\, 6\pi^2 \,}
= \frac{\, p_{\rm sat}^3 \,}{\, 6\pi^2 \,}
+ \frac{\, 1 \,}{\,  2\pi^2 \,} \int^{\infty}_{ p_{\rm sat}  } p^2 \rmd p ~ f_q ( p -p_{\rm sat}; n_B^c) 
\nonumber \\
&&= \frac{\, p_{\rm sat}^3 \,}{\, 6\pi^2 \,}
+ \frac{\, 1 \,}{\,  2\pi^2 \,} \int^{\infty}_{ 0 } (p +p_{\rm sat} )^2 \rmd p ~ f_q ( p; n_B^c) 
\,.
\eeq
Near the saturation, we keep only terms to $O(p_{\rm sat})$. 
\beq
&& \frac{\, P_F^3 \,}{\, 6\pi^2 \,}
\simeq 
 \frac{\, (P_F^c)^3 \,}{\, 6\pi^2 \,}
+ \frac{\, p_{\rm sat} \Lambda_c^2 \,}{\,  2\pi^2 \,} \,,
\eeq
or
\beq
p_{\rm sat} \simeq 
  \frac{\, \pi^2 \big( n_B - n_B^c \big) \,}{\, \Nf \Lambda_c^2 \,} \,,
\eeq
where $\Lambda_c \sim \Lambda $ characterizes the thickness of the distribution at $n_B=n_B^c$,
\beq
\Lambda_c^2 =  \int^{\infty}_{ 0 } \rmd p^2 ~ f_q ( p; n_B^c) \,.
\eeq
In the large $\Nc$, $f_q (p;n_B^c) = \rme^{-\tp^2}$ so that $\Lambda_c \rightarrow \Lambda$.
It is clear that $p_{\rm sat} \rightarrow 0$ as $n_B \rightarrow n_B^c$ from above.
Similarly $\varepsilon$ is continuous before and after the saturation,
\beq
\hspace{-0.5cm}
&& \varepsilon
=  \varepsilon_{\rm sat}
+ \frac{\, \Nc \Nf \,}{\,  \pi^2 \,} \int^{\infty}_{ p_{\rm sat}  } p^2 \rmd p ~ E_q(p) f_q ( p -p_{\rm sat}; n_B^c) 
\,.
\eeq
where $\varepsilon_{\rm sat}$ is the contribution from $p=0$ to $p_{\rm sat}$.
For a small $p_{\rm sat} \sim 0$, we neglect $\varepsilon_{\rm sat}$ and expand the integrand in the second term,
\beq
 \varepsilon
\simeq
 \varepsilon_{c} 
 +
 p_{\rm sat}  \frac{\, \Nc \Nf\,}{\,  \pi^2 \,} 
 \int^{\infty}_{ 0 } \!\! p^2 \rmd p \, E_q(p)
   \bigg(- \frac{\, \partial f_q ( p; n_B^c)  \,}{\, \partial p \,} \bigg) \,,
   \nonumber \\
\eeq
where $\varepsilon_c$ is the energy density at $n_B=n_B^c$. The derivative is $\partial f_q/\partial p^2 \le 0$, so the  $\varepsilon$ approaches $\varepsilon_c$ continuously from above for $p_{\rm sat}\rightarrow 0^+$.
Noting $\partial p_{\rm sat}/\partial n_B \simeq \pi^2 /\Nf \Lambda_c^2$,
the chemical potential just after the saturation is
\beq
\mu_B \simeq  \frac{\, \Nc \,}{\,   \Lambda_c^2 \,}
 \int^{\infty}_{ 0 } \!\! p^2 \rmd p \, E_q(p)
 \bigg(- \frac{\, \partial f_q ( p; n_B^c)  \,}{\, \partial p \,} \bigg) \,,
\eeq
which is compared to the $\mu_B$ before the saturation.

A number of analytic insights are obtained in the large $\Nc$ limit,
where $f_q (p;n_B^c) \rightarrow \rme^{-\tp^2}$ and $\Lambda_c \rightarrow \Lambda$, so that
\beq
\frac{\,  \mu_B^{\rm after} \,}{\Nc}
\, \rightarrow \,
2 \Lambda 
 \int^{\infty}_{ 0 } \!\! \tp^3 \rmd \tp \, \tilde{E}_q(p) \rme^{-\tp^2} \,,
\eeq
where we wrote $\tilde{E} = E/\Lambda$.
The chemical potential before the saturation is given by the baryon mass at large $\Nc$,
\beq
\frac{\, \mu_B^{\rm before} \,}{ \Nc }
\, \rightarrow \,
  \frac{\, 4 \Lambda \,}{\, \sqrt{\pi} \,}  \int_0^\infty \!\! \tp^2 \rmd \tp \, \tilde{E}_q (p) \, \rme^{-\tp^2} = M_B \,.
\eeq
If the saturation takes place in the relativistic regime of quarks, we may expand $E_q \sim p + \cdots$, and find
\beq
\frac{\,  \mu_B^{\rm after} \,}{\, \mu_B^{\rm before} \,}
\, \rightarrow \, 
\frac{\, \sqrt{\pi} \,}{\, 2 \,} \bigg( \frac{\, 3\sqrt{\pi} \,}{4} + \cdots \bigg) \simeq 1.18 + \cdots \,,
\label{eq:muB_rela}
\eeq
where the chemical potential jumps by $\simeq 0.18 M_B$. 
Meanwhile, in the nonrelativistic limit,
\beq
\bigg( \frac{\,  \mu_B^{\rm after} \,}{\, \mu_B^{\rm before} \,} \bigg)_{\rm NR}
\, \rightarrow \, 
1 + \frac{\, \Lambda^2 \,}{\, 4M_q^2 \,} + \cdots  \,,
\label{eq:muB_NR}
\eeq
where the jump in $\mu_B$ is the order of nonrelativistic corrections.

We found that, while $n_B$ and $\varepsilon$ do not contain any jumps, the derivatives do.
Of course, the thermodynamics does not allow jumps in $\mu_B$ and the results being presented must contain something unphysical,
in spite of the seemingly reasonable form of $f_q$ in Eq.(\ref{eq:fq_after_sat}).
But for the moment we proceed further to examine what would happen in this idealized description.

\begin{figure}[tb]
\begin{center}	
\vspace{-0.1cm}
	\includegraphics[width=8.8cm]{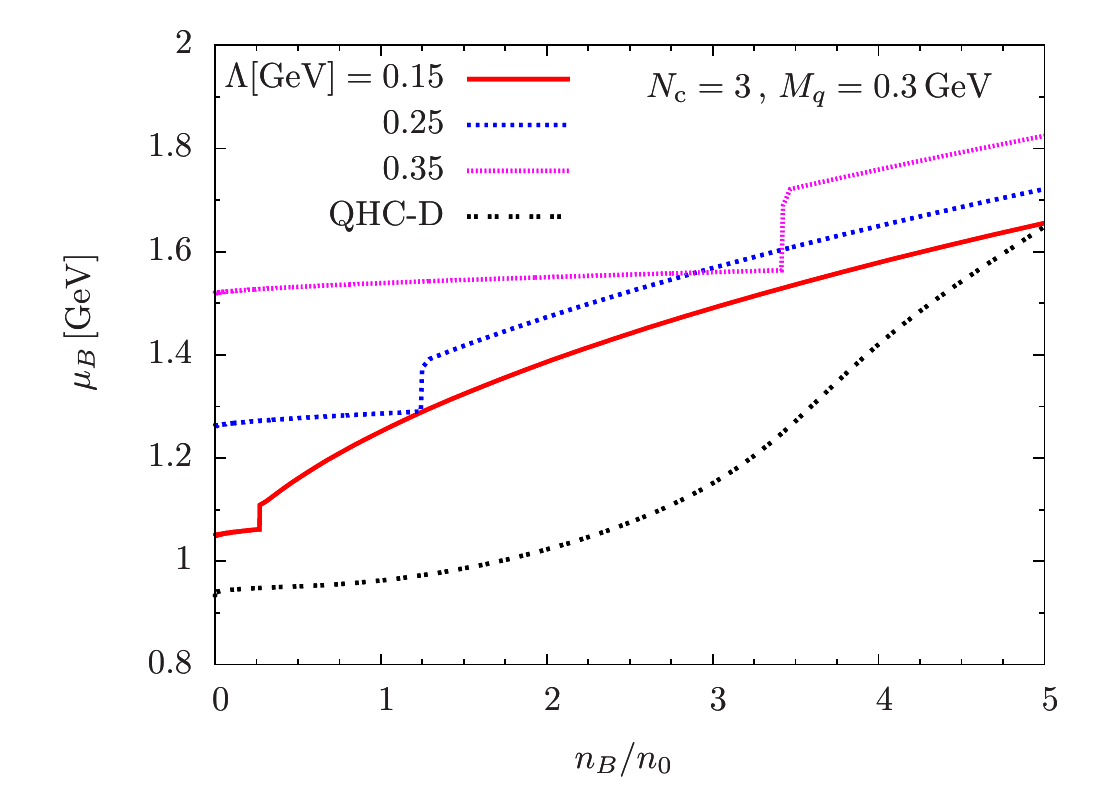}
\caption{ \footnotesize{The $n_B$ dependence of the baryon chemical potential $\mu_B$ for a model Eq.(\ref{eq:fq_after_sat}) without interactions. The results of $\Nc=3$, $M_q=0.3$ GeV, and $\Lambda=0.15, 0.25, 0.35$ GeV are shown.
QHC-D is also plotted as a reference.
		} }
	\vspace{-0.5cm}
\label{fig:FIG_eos_lam0.25_Mq0.30_CM0.0_Vc0.0_gv0.0_muB-nB}		
\end{center}
\end{figure}

Figure \ref{fig:FIG_eos_lam0.25_Mq0.30_CM0.0_Vc0.0_gv0.0_muB-nB} shows $\mu_B$  as a function of $n_B/n_0$ where
we chose $\Nc=3$, $M_q=0.3$ GeV, and $\Lambda=0.25$ GeV, for which $M_B\simeq 1.26$ GeV,
and the jump in $\mu_B$ associated with the saturation is $\simeq 0.1$ GeV.
It is clear that the $\mu_B$ and $\varepsilon/n_B$ after the saturation grow much faster than the behavior before the saturation.
For comparisons, we also show the results for the Quark-Hadron-Crossover (QHC)19-D EoS \cite{Baym:2019iky} as an example of EoS consistent with NS observations;
for the QHC19-D, $M_{\rm max}\simeq 2.28M_\odot$, $R_{1.4} \simeq 11.6$ km, and $R_{2.08}\simeq 11.5$ km.
(In Sec.\ref{sec:interactions}, we make comparisons again after adjusting the baryon mass.)

\begin{figure}[tb]
\begin{center}	
\vspace{-0.1cm}
	\includegraphics[width=8.8cm]{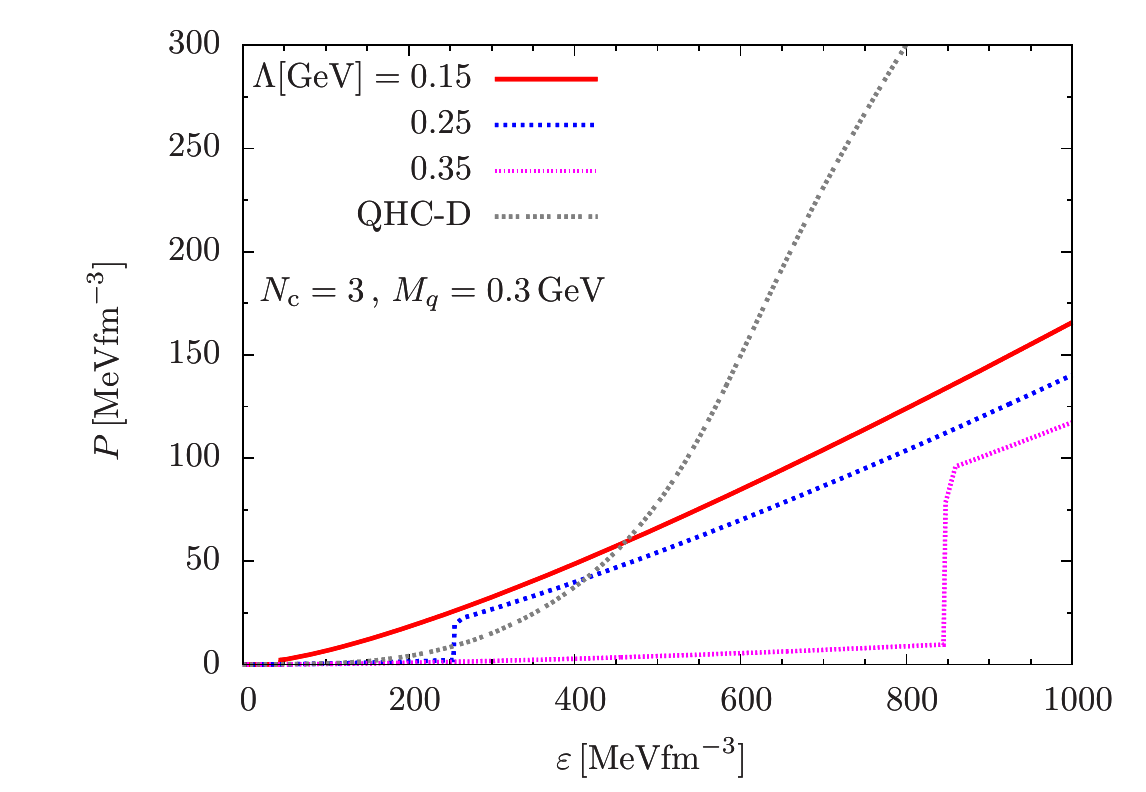}
\caption{ \footnotesize{The $\calP$ vs $\varepsilon$ for the setup same as Fig.\ref{fig:FIG_eos_lam0.25_Mq0.30_CM0.0_Vc0.0_gv0.0_muB-nB}.
		} }
	\vspace{-0.5cm}
\label{fig:FIG_eos_lam0.25_Mq0.30_CM0.0_Vc0.0_gv0.0_e-P}		
\end{center}
\end{figure}

These jumps in $\mu_B$ result in the discontinuities in the corresponding pressure $\calP$ through the thermodynamic relation $\calP=\mu_B n_N - \varepsilon$, where $n_B$ and $\varepsilon$ are continuous but $\mu_B$ are not. 
Now, the pressure just after the saturation is
\beq
\hspace{-0.5cm}
\calP\, \simeq \, \mu_B^{\rm after} n_B -\varepsilon \simeq \big( \mu_B^{\rm after} - \mu_B^{\rm before} \big) n_B^c  \sim \Nc \Lambda^4
 \,,
\eeq
which is much bigger than the pressure of an ideal baryon gas, $\calP \sim n_B^{5/3}/M_B \sim (P_F/\Lambda)^3/\Nc $.
Accordingly the squared speed of sound $c_s^2 = \partial \calP/\partial \varepsilon$ diverges to $+\infty$ at the saturation.
The $\calP$ vs $\varepsilon$ 
for the setup, same as Fig.\ref{fig:FIG_eos_lam0.25_Mq0.30_CM0.0_Vc0.0_gv0.0_muB-nB}, is shown in Fig.\ref{fig:FIG_eos_lam0.25_Mq0.30_CM0.0_Vc0.0_gv0.0_e-P}.

\subsection{Smoothing out the discontinuities}
\label{sec:model_saturation}

\begin{figure}[tb]
\begin{center}	
\vspace{-0.4cm}
	\includegraphics[width=8.8cm]{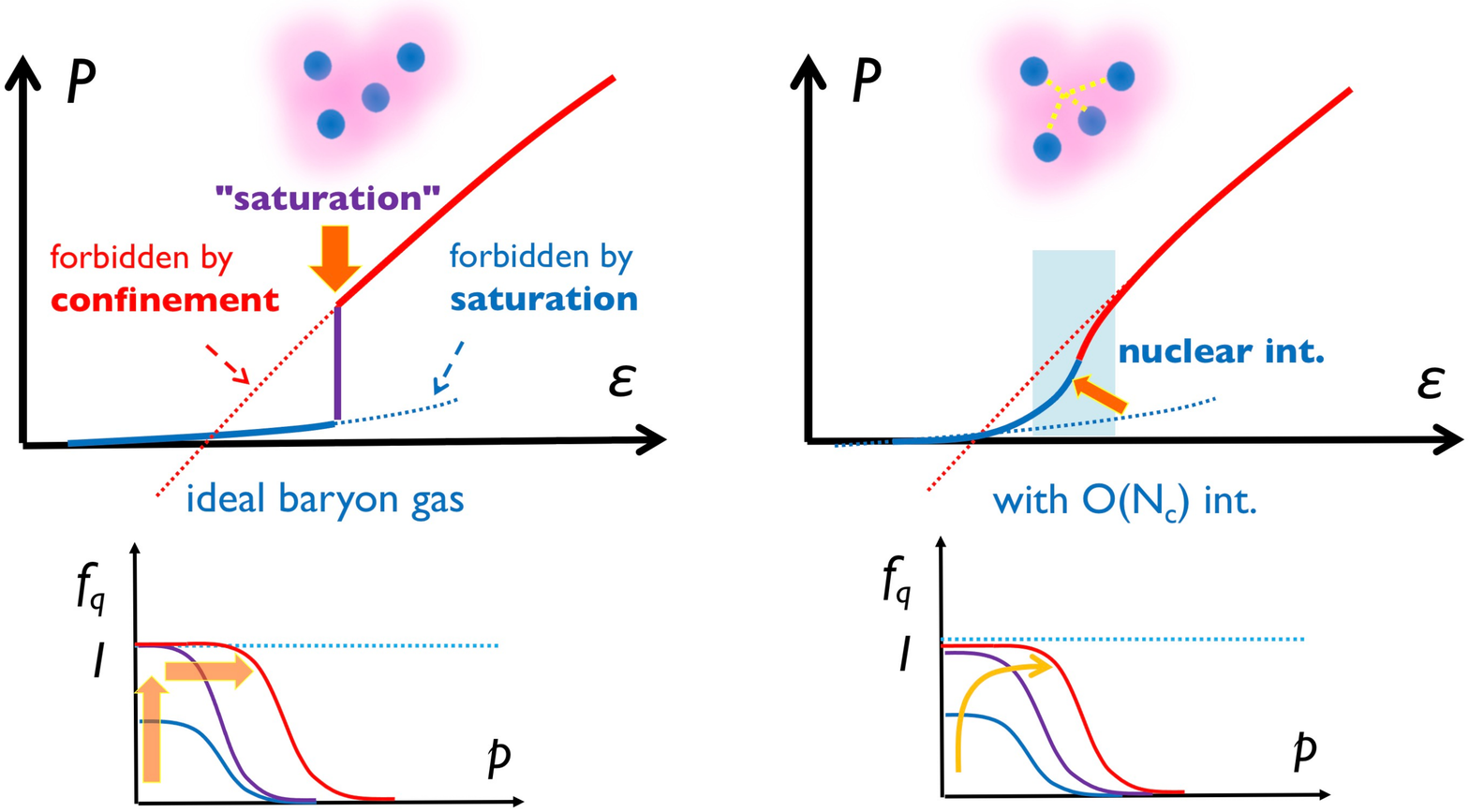} 
	\vspace{-0.8cm}
\caption{ \footnotesize{
Smoothing out the artifact: The use of ideal baryon gas description to the quark saturation point leads to a discontinuous jump in $\calP$ (left panel).
With baryon interactions mediated by quark exchanges, the precursory behavior should appear before reaching the quark saturation point (right panel). 
In the lower panels, the evolution of $f_q$, which changes the direction from the vertical to horizontal direction at the quark saturation point, takes place more smoothly with baryon interactions.
		} }
	\vspace{-0.5cm}
\label{fig:int_smearing}		
\end{center}
\end{figure}

\begin{figure}[tb]
\begin{center}	
\vspace{-0.1cm}
	\includegraphics[width=8.8cm]{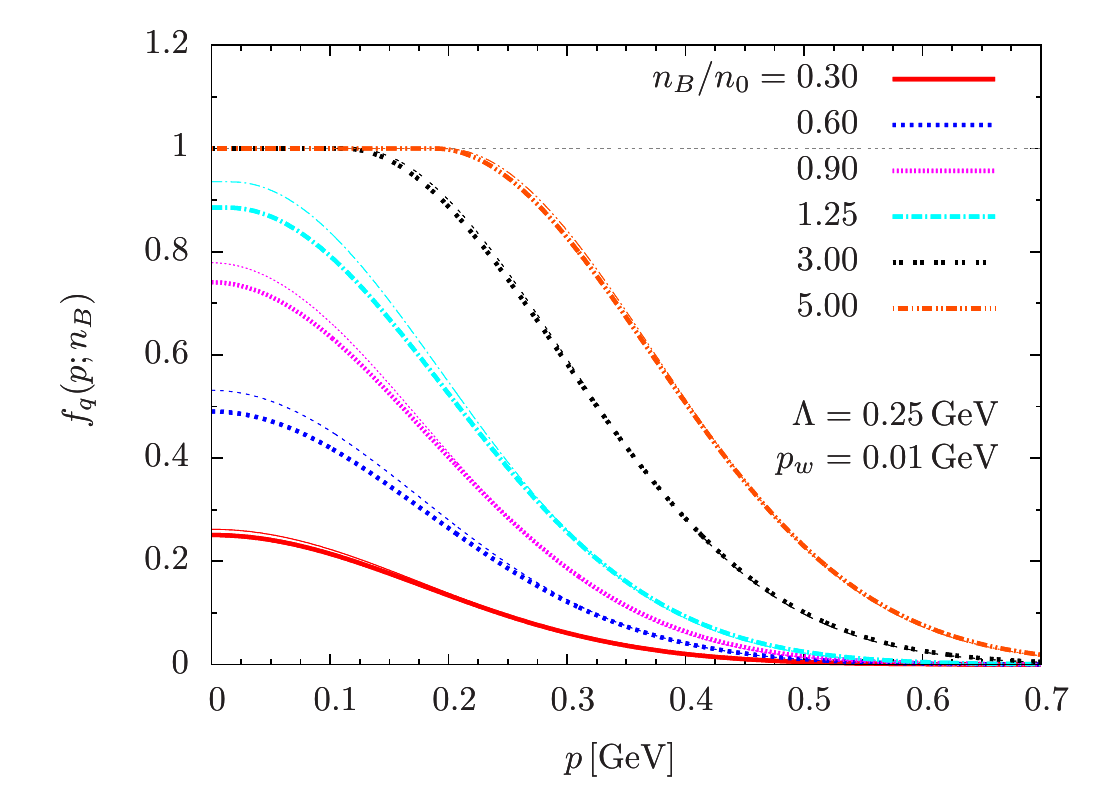} 
\caption{ \footnotesize{The quark occupation probability for $f^{\rm sat}_q(p;n_B)$ in Eq.(\ref{eq:fq_sat_smear}), for various $n_B/n_0$. We took $\Lambda=0.25$ GeV, $p_w=0.01$ GeV, and $\Nc=3$.
	The thin lines are the leading order of the $1/\Nc$ expansion. 
		} }
	\vspace{-0.5cm}
\label{fig:fq_for_pbf_lam_sat}		
\end{center}
\end{figure}

The discontinuities are the artifacts which are presumably associated with our use of the ideal baryon gas picture for baryons just before the quark saturation (Fig. \ref{fig:int_smearing}).
The validity of the ideal gas picture should break down before reaching the quark saturation point.
The baryon interactions are mediated by the meson exchanges or quark exchanges, so near the quark saturation point those interactions should grow up.
In realistic nuclear EoS, nuclear interactions of $O(\Nc)$ are important, and the pressure increases faster than in the ideal baryon gas.
Thus, with interactions the pressure does not suddenly jump up, but there should be a precursory behavior as the system approaches the quark saturation regime.

With these considerations in mind, we introduce an {\it ad hoc} procedure to smooth out the discontinuities. 
We revise the previous model of quark saturation slightly, by multiplying a smearing function,
\beq
&& f_q^{\rm sat} (p; p_{\rm sat})
=
\tanh( p_{\rm sat}/p_w ) \nonumber \\
&&\hspace{0.5cm}
 \times 
\big[\, \theta(p_{\rm sat}-p) + \theta( p- p_{\rm sat} ) \, \rme^{- (\tp-\tp_{\rm sat})^2 } \big]\,,
\label{eq:fq_sat_smear}
\eeq
where $p_w$ is a regulator which should be taken to be very small, $p_w \ll \Lambda$. For $p_{\rm sat} \ll p_w$, $p_{\rm sat}$ in the step function and the Gaussian factor is negligible, so we can regard
\beq
f_q^{\rm sat} \simeq \frac{p_{\rm sat} }{p_w} \, \rme^{- \tp^2 } \, \sim \, \frac{p_{\rm sat} }{p_w} \, f_q (p; n_B = 0) \,,
\eeq
in computations of thermodynamic quantities.
Technically, the factor $p_{\rm sat}/p_w \ll 1$ plays the role similar to the factor $\sim \calN P_F^3 \sim n_B/n_B^c$ in Eq.(\ref{eq:fq_pF}). There the shape of $f_q$ was fixed but its magnitude increases until the saturation takes place.
When $p_{\rm sat} \sim p_w$, the model goes back to the model in Eq.(\ref{eq:fq_after_sat}) modulo $\tP_F \sim 1/\Nc$ corrections.
The $n_B$ dependence of $f_q^{\rm sat}$ for $p_w=0.01$ GeV is shown in Fig.\ref{fig:fq_for_pbf_lam_sat}.
The qualitative behaviors are very similar to Fig.\ref{fig:fq_for_pbf_lam}
[the major difference comes from our neglect of $\tP_F$ corrections in Eq.(\ref{eq:fq_sat_smear}) compared to Eq.(\ref{eq:fq_after_sat})].
	
\begin{figure}[tb]
\begin{center}	
\vspace{-0.1cm}
	\includegraphics[width=8.8cm]{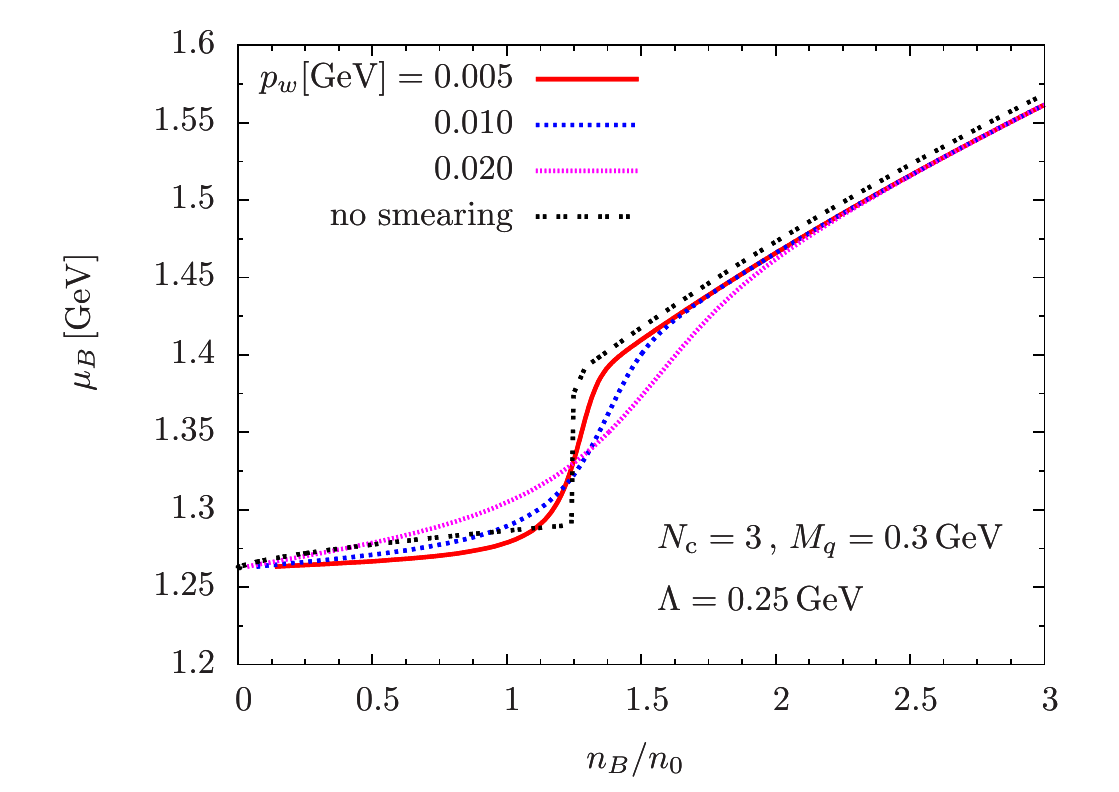}
\caption{ \footnotesize{ The $\mu_B$ vs $n_B/n_0$ for the $p_w=0.005, 0.010, 0.020$ GeV. We took $\Lambda=0.25$ GeV, $M_q=0.3$ GeV, and $\Nc=3$.
		} }
	\vspace{-0.5cm}
\label{fig:smear_muB-nB}		
\end{center}
\end{figure}

\begin{figure}[tb]
\begin{center}	
\vspace{-0.1cm}
	\includegraphics[width=8.8cm]{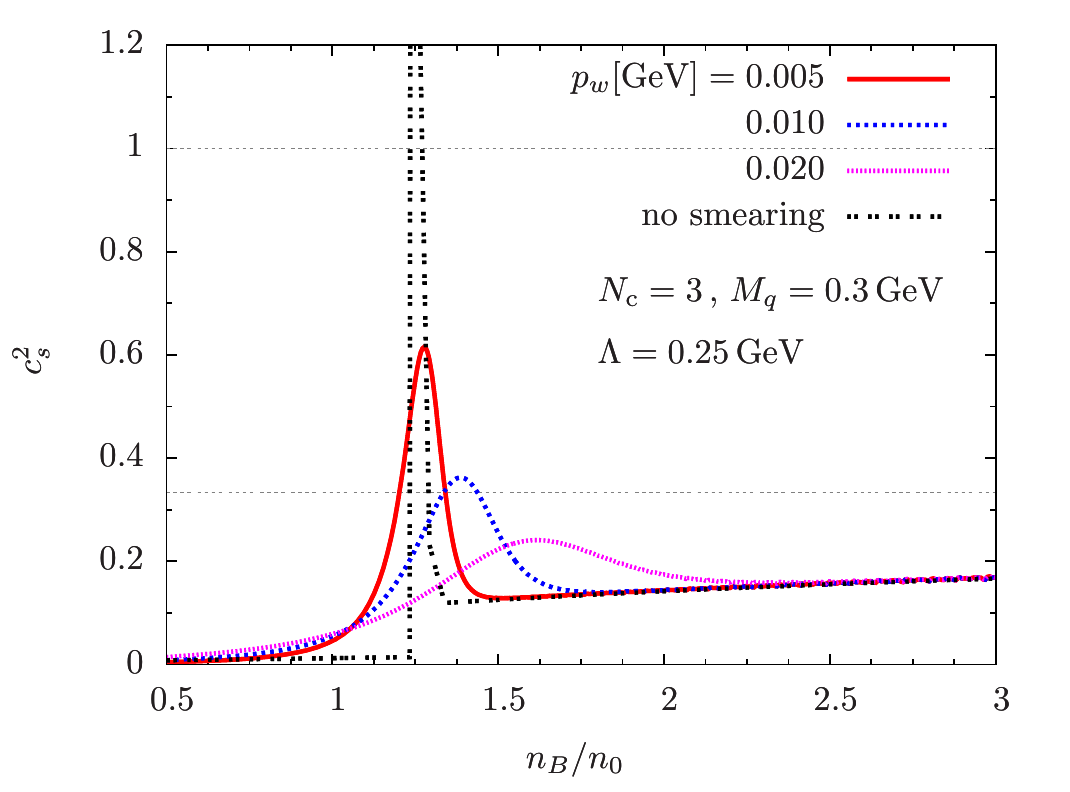}
\caption{ \footnotesize{ The squared speed of sound $c_s^2$ vs $n_B/n_0$ for the parameter set same as Fig.\ref{fig:smear_muB-nB}.	
		} }
	\vspace{-0.5cm}
\label{fig:smear_nb-cs2}		
\end{center}
\end{figure}

Now, we use $f_q^{\rm sat}$ to examine several thermodynamic quantities.
Shown in Fig.\ref{fig:smear_muB-nB} is $\mu_B$ as a function of $n_B$.
Working with a model with the same form from low to high density, the artificial discontinuities found in the previous treatment are smoothed out, and $\mu_B$ is now continuous, as it should. 
Accordingly, the (squared) speed of sound $c_s^2$ is now regulated and well defined, as shown in Fig.\ref{fig:smear_nb-cs2}.	
The $c_s^2$ has a peak around the density where the saturation effects become important, and it exceeds the conformal value $1/3$ for a sufficiently small $p_w$.
The degree of the smearing is rather sensitive to the value of $p_w$, and a too small $p_w$ tends to violate the causality constraint $c_s^2 \le 1$.
The sharpness of the peak should be dynamically determined by the interplay between baryon and quark dynamics.
We choose $p_w=0.01-0.02$ GeV unless otherwise stated.

We emphasize that the existence of the peak was not driven by nuclear forces, but by the quark Pauli blocking effects.
Including baryon interactions tames the peak, rather than enhancing it;
stronger baryonic interactions before the quark saturation blur the peak structure, by reducing the contrast between baryonic and quark matter pressures.
Taking the picture that baryon interactions are mediated by quark exchanges,
baryonic matter with stronger interactions should look more similar to quark matter\footnote{
This consideration is in conflict with some EoS models with first order hadron-quark phase transitions.
Such models assume strong repulsions in nuclear matter to describe very stiff nuclear EoS; 
as a consequence the quark EoS can remain stiff even after first order phase transitions, being consistent with the $2M_\odot$ constraints.
In this scenario nuclear matter with stronger interactions differs more from quark matter.
In contrast, our picture in this paper takes the view that baryonic matter with stronger interactions is closer to quark matter.
 }.
 This viewpoint motivates a three-window modeling of dense matter from nuclear to quark matter domains \cite{Masuda:2012kf,Masuda:2012ed,Masuda:2015kha,Kojo:2014rca,Baym:2017whm,Baym:2019iky,Ma:2019ery,Ayriyan:2021prr}, 
 and more microscopic considerations based on mode-by-mode percolation  \cite{Fukushima:2020cmk}.

\section{Baryons after saturation}
\label{sec:occ_baryon}

As we have seen, the saturation of low momentum quark states induces a rapid increase in the pressure. 
Such changes are difficult to imagine from purely baryonic descriptions, and we are interested in how the corresponding occupation probability of baryon states ($\calB$) looks.
Through attempts to understand $\calB$, we also discuss how to obtain expressions similar to the MR model for quarkyonic matter EoS \cite{McLerran:2018hbz}.

Regarding (for a given $n_B$) $f_q$ and $\calB$ as vectors with indices $\vp$ and $\vP_B$, respectively, they are related through a matrix $Q_{\rm in}$ as 
\beq
\vec{f}_q = Q_{\rm in} \vec{\calB} \,,
\eeq
where $0 \le \calB (P_B;n_B) \le 1$ for any $P_B$ and $n_B$.  
The equation is linear and in principle one can take inversion to determine $\calB$ for a given $f_q$.
But in practice this method does not work well.
Another strategy is to prepare some model $\calB_{\alpha}$ for $\calB$ with parameters $\vec{\alpha} = (\alpha_1, \alpha_2,\cdots )$,
and minimize a function ``energy'' functional
\beq
\hspace{-0.4cm}
\calH (\vec{\alpha}) 
= \big( \vec{f}_q - Q_{\rm in} \vec{\calB}_\alpha \big)^T \big( \vec{f}_q - Q_{\rm in} \vec{\calB}_\alpha \big)
+ \calI_{\rm cost} ( \calB_\alpha ) \,,
\eeq
where $\calI_{\rm cost}$ is some functional which gives the energetic penalty 
if, for instance, $\calB_\alpha$ violates the condition $0 \le \calB (P_B;n_B) \le 1$.
For this strategy to work, we need a good guess for the form of $\calB$.

\begin{figure}[tb]
\begin{center}	
\vspace{-1.3cm}
	\includegraphics[width=8.5cm]{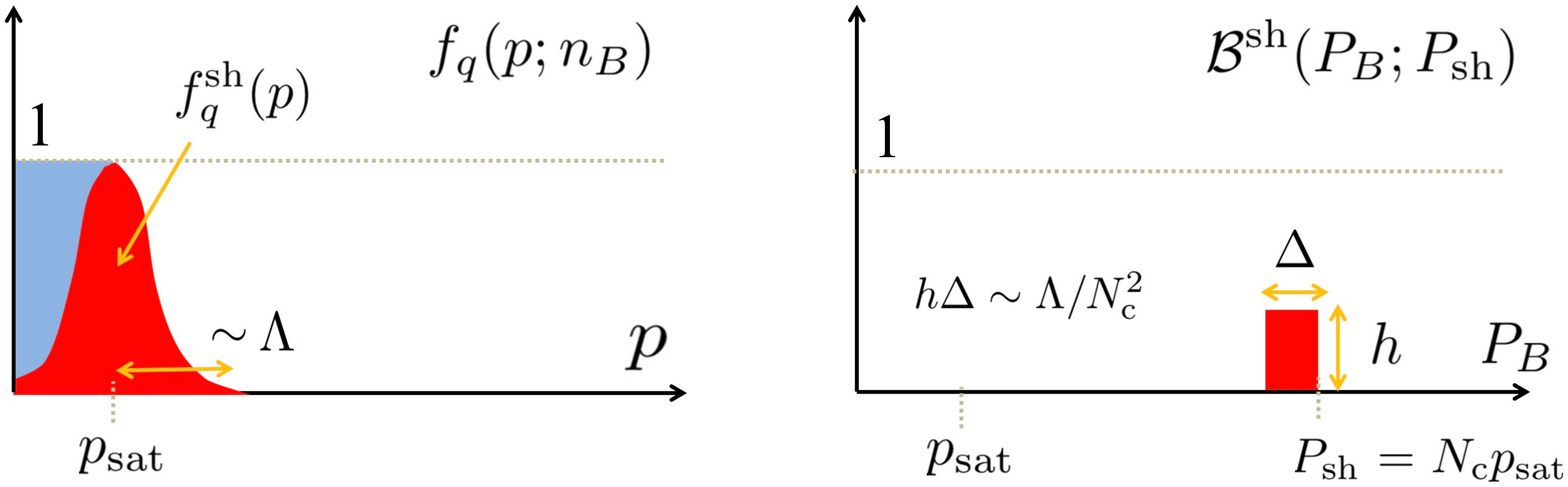}
		\vspace{-1.5cm}
\caption{ \footnotesize{ 
A model $\calB^{\rm sh}$ for a baryon momentum distribution after the saturation takes place. 
The $\calB^{\rm sh}$ convoluted with $Q_{\rm in}$ leads to the quark momentum distribution $f_q^{\rm sh}$ around $p_{\rm sat}$.
For the ``baryon shell'' of $\calB^{\rm sh}$, we found $h\Delta \sim 1/\Nc^2$ for $P_{\rm sh} \sim \Nc \Lambda$.
If we fix $h=1$, the thickness $\Delta$ gets narrower as in the MR model. 
 		} }
	\vspace{-0.5cm}
\label{fig:Bsh_to_fq}		
\end{center}
\end{figure}

For this purpose, we first assume baryons with momenta $ P_B \gtrsim \Lambda$ and discuss which form of $f_q$ should follow.
We consider the form
\beq
\calB^{\rm sh} (P_B; P_{\rm sh}) = h \theta( P_{\rm sh}  - P_B) \theta( P_B - P_{\rm sh} -\Delta) \,,
\label{eq:shell_calB}
\eeq
where ``sh'' is the abbreviation of ``shell'', and $h$ is the overall size in the occupation probability ($0 \le h \le 1$) for states with momenta $P_{\rm sh} - \Delta (\ge 0)$ to $P_{\rm sh}$.
Below, to simplify calculations, we assume $\Delta$ to be small, $\Delta \ll P_{\rm sh}$.
Integrating the baryon momentum distribution, we get the contribution to $f_q$ as
\beq
 \hspace{-0.2cm}
  f_q^{\rm sh} (p)  
\simeq 
\frac{\, h \Delta \,}{ \Lambda }\, 
 \frac{\, \Nc^3 \,}{\,  \sqrt{\pi} \,} 
\frac{\, \tP_{\rm sh} \,}{\,  \tp \,}\,
 \rme^{-\tp^2-\tP_{\rm sh}^2 }
\, \big(\, \rme^{2\tp \tP_{\rm sh} }  - \rme^{- 2 \tp \tP_{\rm sh} } \, \big)
\,,
\eeq
where we set $\tP_{\rm sh} = P_{\rm sh}/\Nc \Lambda$.
For small $\tp \tP_{\rm sh}$,
\beq
 f_q^{\rm sh} (p)  
\sim
4 
\frac{\, h \Delta \,}{ \Lambda }\, 
\,  \frac{\, \Nc^2 \,}{\,  \sqrt{\pi} \,} \, \rme^{-\tp^2-\tP_{\rm sh}^2 }\,
 \tP_{\rm sh}^2 \big(\, 1 + O(\tp^2 \tP^2_{\rm sh}) \,\big)
\,,
\label{eq:f_q_sh_small_p}
\eeq
and for large $\tp \tP_{\rm sh}$,
\beq
 \hspace{-0.2cm}
  f_q^{\rm sh} (p)  
\sim
\frac{\, h \Delta \,}{ \Lambda }\, 
 \frac{\, \Nc^2 \,}{\,  \sqrt{\pi} \,} 
\frac{\, \tP_{\rm sh} \,}{\,  \tp \,}\,
 \rme^{- (\tp - \tP_{\rm sh} )^2 }
\,,
\label{eq:f_q_sh_large_p}
\eeq
where the maximum of $f_q$ appears at $\tp \sim \tP_{\rm sh}$.

Below, we examine these asymptotic behaviors in some details.
We can see a number of the remarkable features in the large $\Nc$ limit.
For $P_{\rm sh} \sim \Lambda$ or $\tP_{\rm sh} \sim 1/\Nc$,
we use Eq.(\ref{eq:f_q_sh_small_p}) to obtain
\beq
f_q^{\rm sh} (p) \, \sim \, 
\frac{\, h \Delta \,}{ \Lambda }\, 
 \, \rme^{-\tp^2} \,,
\label{eq:scaling_small_p}
\eeq
where the condition $f_q \le 1$ demands $h\Delta \sim \Lambda$.

For $P_{\rm sh} \sim \Nc \Lambda$ or $\tP_{\rm sh} \sim 1$,
we use Eq.(\ref{eq:f_q_sh_large_p}) to obtain
\beq
 \hspace{-0.2cm}
  f_q^{\rm sh} (p)  
\, \sim \, 
\frac{\, h \Delta \,}{ \Lambda }\, 
\Nc^2 \,
 \rme^{- (\tp - \tP_{\rm sh} )^2 }
\,,
\label{eq:scaling_large_p}
\eeq
where we set $\tP_{\rm sat}/\tp \sim 1$ as the Gaussian part has a peak at $\tp = \tP_{\rm sh}$.
In this regime the condition $f_q \le 1$ demands $h\Delta \sim \Lambda /\Nc^2$.

Now we consider the forms of $h$ that are compatible with the scaling behaviors in Eqs.(\ref{eq:scaling_small_p}) and (\ref{eq:scaling_large_p}). 
For example, one can take
\beq
[ h \Delta ] ( P_{\rm sh} ) 
\sim 
c_0 \Lambda \bigg(  \frac{\, \Lambda^2 \,}{\, P_{\rm sh}^2 \,} + \frac{\, c_1 \,}{\, \Nc \,} \frac{\Lambda}{\, P_{\rm sh} } + \frac{c_2}{\, \Nc^2 } \bigg) \,,
\eeq
where $c_0, c_1,$ and $c_2$ are constants of $O(1)$. 
For $P_{\rm sh} \sim \Lambda$, the first term dominates. For $P_{\rm sh} \sim \Nc \Lambda$, all these three terms can be comparable.
In what follows, baryon states with large momenta $P_{\rm sh}$ are compatible with the condition $f_q \le 1$ only if
those states are occupied with the small probability density, i.e., either $h$ or $\Delta$ becomes small for large $P_{\rm sh}$.
In particular baryon states with momenta $\sim \Nc \Lambda$ are occupied with a small but nonzero probability $h\Delta \sim 1/\Nc^2$.
In the Appendix, we further explore the phase space density in baryonic descriptions.

Now we notice that the scaling $f_q^{\rm sh} \sim \rme^{-(\tp- \tP_{\rm sh})^2 } $ is similar to the behavior of 
$f_q( p - p_{\rm sat} ; n_B^c) \sim \rme^{-(\tp-\tp_{\rm sat})^2 }$ in Eq.(\ref{eq:fq_after_sat}) for the partially occupied quark states beyond the saturated states. 
This suggests that $\tP_{\rm sh} \simeq \tp_{\rm sat}$ or $P_{\rm sh} \simeq \Nc p_{\rm sat}$, and the approximate relation
\beq
\hspace{-0.5cm} 
f_q( p - p_{\rm sat} ; n_B^c) 
\sim \int_{\vP_B} Q_{\rm in} \, \calB^{\rm sh} (P_B; P_{\rm sh} \simeq \Nc p_{\rm sat}) \,.
\eeq
Thus the shell form Eq.(\ref{eq:shell_calB}) turns out to be a good candidate for the $p>p_{\rm sat}$ part of the quark distribution $f_q$ postulated in Eq.(\ref{eq:fq_after_sat}).
The schematic picture is given in Fig.\ref{fig:Bsh_to_fq}.

Substituting this form into Eq.(\ref{eq:fq_after_sat}), and then integrating quark momenta $p$, we reach a model similar to the model of 
McLerran and Reddy \cite{McLerran:2018hbz}.\footnote{Our presentation here is slightly different from the paper of McLerran and Reddy \cite{McLerran:2018hbz} where the concept of the occupation probability for baryons is not used, so $h = 1$ from the very beginning, while they assumed the form of $\Delta$ to be
\beq
\Delta = \frac{\, \Lambda^3 \,}{\, P_{\rm sh}^2 \,} + \frac{\, \kappa \Lambda \,}{\, \Nc^2 \,} \,.
\eeq
The second term plays important roles to regulate $\mu_B$ and the speed of sound.
}
With $P_{\rm sh}=\Nc p_{\rm sat}$, the baryon number and energy densities per flavor are
\beq
\hspace{-0.5cm} 
\frac{\, n_B \,}{\, \Nf \,} 
&\simeq& 
  \frac{\, h \,}{\, \pi^2 \,} \int_{ P_{\rm sh} - \Delta }^{ P_{\rm sh} } \!\! \rmd P_B P_B^2
+  \frac{\, p_{\rm sat}^3 \,}{\, 3\pi^2 \,}
\nonumber \\
\frac{\, \varepsilon \,}{\, \Nf \,}
&\simeq &
 \frac{\, h \,}{\, \pi^2 \,} \int_{ P_{\rm sh} - \Delta }^{P_{\rm sh}  } \!\! \rmd P_B \, P_B^2 E_B 
+ \Nc \int_0^{p_{\rm sat}} \!\! \frac{\, \rmd p \, p^2 \,}{\, \pi^2 \,} \, E_q (p)\,,
\nonumber \\
\eeq
where we have used the relations $\int_{\vp} Q_{\rm in} = 1$ and $E_B (P_B) = \Nc \int_{\vp} Q_{\rm in} (\vp,\vP_B) E_q(p)$, and neglected possible double counting  in $f_q$ at $p\lesssim p_{\rm sat}$
which should be minor effects unless $p_{\rm sat}$ is very large.
We also note that
\beq
\mu_B 
&=& \frac{\, \partial P_{\rm sh} \,}{\, \partial n_B \,}  \frac{\, \partial \varepsilon \,} {\, \partial P_{\rm sh} \,} \,,
\nonumber \\
&\simeq &
 \frac{\, P_{\rm sh}^2 E_B(P_{\rm sh}) - ( P_{\rm sh} -\Delta)^2 E_B(P_{\rm sh}-\Delta)  \,}
	{\, P_{\rm sh}^2 - ( P_{\rm sh} -\Delta)^2  \,}
\eeq
where the contributions from the saturated quark states are suppressed by an extra factor of $1/\Nc$.
When $P_{\rm sh} \sim \Lambda $, baryons are nonrelativistic, 
$E_B(P_{\rm sh}-\Delta) \simeq E_B(P_{\rm sh})$ and $\mu_B -M_B \simeq E_B(P_{\rm sh}) -M_B \sim 1/\Nc$.
As $P_{\rm sh}$ approaches $\Nc \Lambda ( \gg \Delta )$, we find $\mu_B -M_B  \simeq E_B (P_{\rm sh}) -M_B + (P_{\rm sh}/2) \partial  E_B/\partial P_{\rm sh} \sim P_{\rm sh}^2/E(P_{\rm sh}) \sim \Nc \Lambda$,
as in usual relativistic expressions.

What is remarkable is that the relativistic regime is reached already for $n_B\sim \Lambda^3$.
For $P_{\rm sh} \sim \Nc \Lambda$, we obtain  
\beq
n_B 
&\simeq&
 \frac{ h }{\, \pi^2 \,} \big( P_{\rm sh}^3 - ( P_{\rm sh} - \Delta )^3 \big)
\sim h \Delta P_{\rm sh}^2
 \nonumber \\
&\simeq & 
c_0 \Lambda^3 + c_1 \Lambda^2 \frac{\, P_{\rm sh} \,}{\, \Nc \,} + c_2 \Lambda \bigg( \frac{\, P_{\rm sh} \,}{\, \Nc \,} \bigg)^2
 \,,
\eeq
which is $\sim \Lambda^3$. For small $P_{\rm sh}$, the first term dominates, while the second and third terms slowly grow as $P_{\rm sh} $ becomes $\sim \Nc\Lambda$.
This result drastically differs from naive expectation $n_B \sim (\Nc \Lambda)^3$.
In ordinary baryonic descriptions at $n_B\sim \Lambda^3$, baryons approach the relativistic regime by interactions of $O(\Nc)$,  
but here we found that relativistic baryons can also emerge just due to the constraints on baryons as composite particles.
This reproduces the remarkable findings by McLerran and Reddy \cite{McLerran:2018hbz}.

\section{Quantum numbers}
\label{sec:quantum_numbers}

\begin{figure}[tb]
\begin{center}	
\vspace{-1.cm}
	\includegraphics[width=8.5cm]{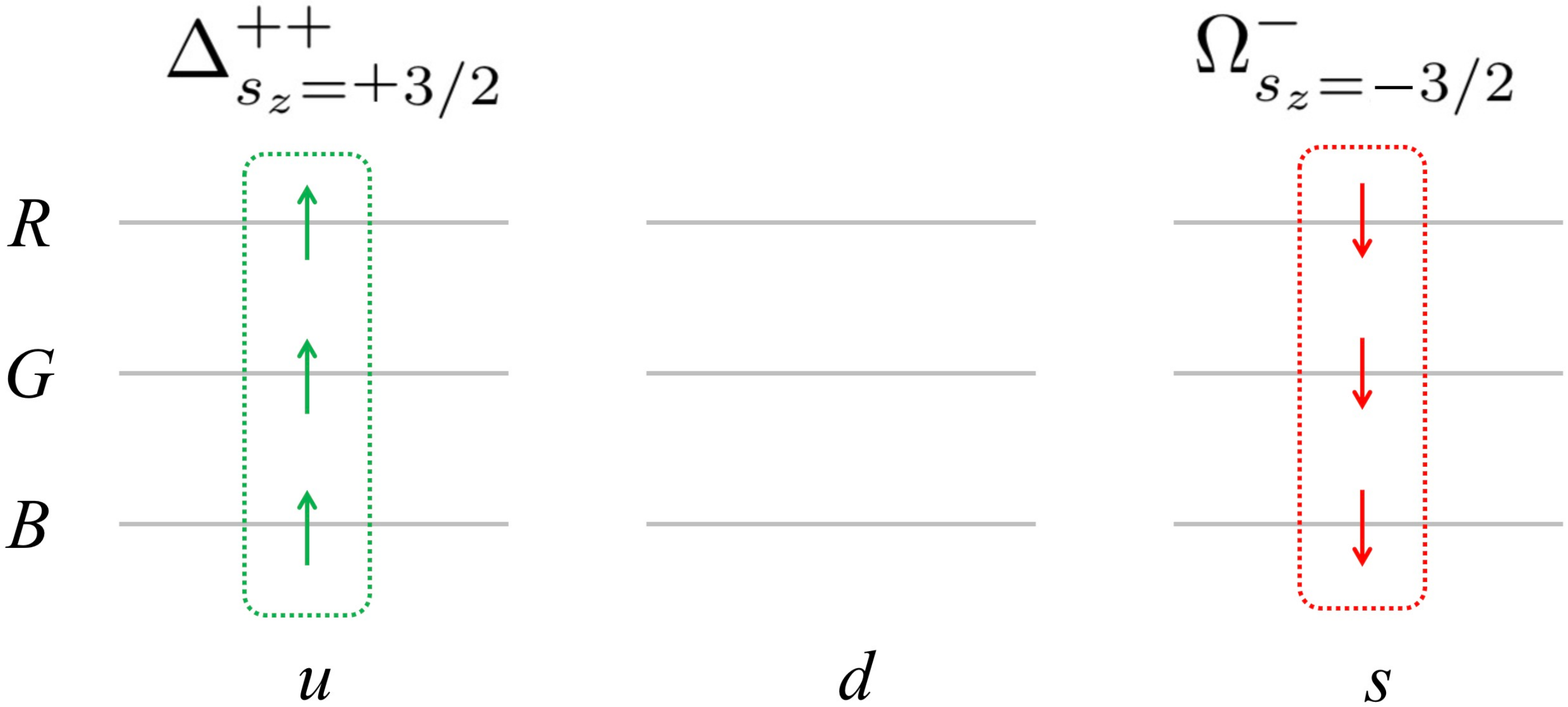}
		\vspace{-1.2cm}
\caption{ \footnotesize{ 
Examples of quantum numbers for baryons where
$\Delta^{++}_{s_z=+3/2}$ and $\Omega^-_{s_z=-3/2}$ are shown.
The former (latter) fills the $u\up$ ($d\down$) states for all colors $RGB$, while leaving the other color-flavor-spin states empty.
For a given space wave function for quarks, six baryon states completely fill color-flavor-spin states.
		} }
	\vspace{-0.2cm}
\label{fig:baryon_quantum_number}		
\end{center}
\end{figure}

We discuss the quantum numbers of baryons such as colors, flavors, and spins.
The question is how baryons saturate these quantum numbers for quark states.
In previous sections, we implicitly assumed that the quark states for $\Nc$ colors, $\Nf$ flavors, and two spins
are saturated by putting $\Nf$ species of baryons with two spins.
In this section we explain why this description should be valid.
Specifically, we consider the $\Nc=\Nf=3$ cases to utilize the terminology common in hadron spectroscopy.
We also ignore the mass difference among up, down and strange quarks.

The argument is simplified by assuming $SU(2\Nf)$ symmetry in which there are no energy differences associated with spins and flavors.
In particular nucleons, $\Delta$, $\Omega$, so on, are all energetically degenerate.
It is useful to focus on the following states,
\beq
\Delta^{++}_{s_z = \pm 3/2} &=& [u_R \! \up u_G \! \up u_B \! \up] \,, ~~ [u_R \! \down u_G \! \down u_B \! \down] \,,  
\nonumber \\
\Delta^{-}_{s_z = \pm 3/2} &=& [d_R \! \up d_G \! \up d_B \! \up] \,, ~~ [d_R \! \down d_G \! \down d_B \! \down] \,,
\nonumber \\
\Omega^{-}_{ s_z = \pm 3/2 } &=& [s_R \! \up s_G \! \up s_B \! \up] \,, ~~ [s_R \! \down s_G \! \down s_B \! \down] \,,
\label{eq:baryon_spin_flavor}
\eeq
where quark states are totally antisymmetrized. 

While there are many more S-wave baryons with spin or flavor excitations (such as baryon octet or decuplet),
for given spatial wave functions we can saturate quark states using only the $2\Nf(=6)$ states listed in Eq.(\ref{eq:baryon_spin_flavor}).
For example (Fig.\ref{fig:baryon_quantum_number}), by putting $\Delta^{++}_{s_z=3/2}$, colors $RGB$ for up quarks with spins aligned in $\up$ directions are filled at once.
Therefore the number of baryon species we need is the same as the number of quark species,
as we have assumed in the previous sections.
The arguments are applicable to any $\Nc$.

This discussion also suggests that, once baryons with a specific spin-flavor quanta form their Fermi sea, the other baryons cannot freely enter the system due to the Pauli blocking at quark level.
For instance, when $\Delta^{++}_{s_z=3/2}$ and $\Delta^{++}_{s_z=-3/2}$ form the large Fermi sea of up quarks, one cannot put nucleons ($uud$ or $ddu$) at low energy,
but must place them on top of the saturated Fermi sea.
This viewpoint should be important when we describe hyperons entering dense nuclear matter.
Also, this consideration should affect the treatment in the self-energy processes for protons and neutrons, as virtual states are blocked by preoccupied states.

More realistically we need to discuss the cases where protons and neutrons appear as the lowest energy states.
The treatments of quantum numbers are more involved than the idealized $(\Delta^{++}, \Delta^-, \Omega^-)$ baryon bases used here.
We leave such realistic cases for our future work.

\section{Interactions}
\label{sec:interactions}

\begin{figure}[tb]
\begin{center}	
\vspace{-1.2cm}
	\includegraphics[width=8.8cm]{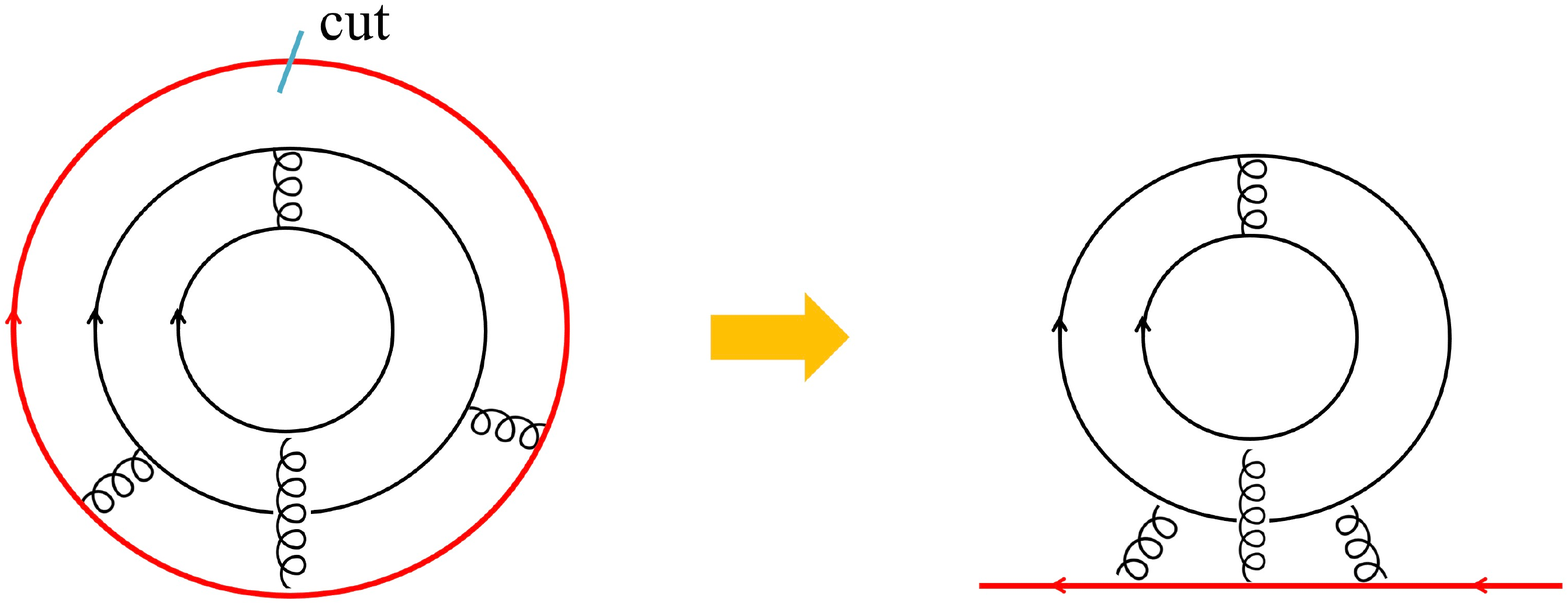}
\vspace{-1.8cm}
\caption{ \footnotesize{
A self-energy graph for a quark in a baryon. The gluon exchanges happen in the color antisymmetric channels.
		} }
		\vspace{-0.5cm}
\label{fig:cut_in_B}		
\end{center}
\end{figure}
\begin{figure}[tb]
\begin{center}	
\vspace{-0.7cm}
	\includegraphics[width=8.8cm]{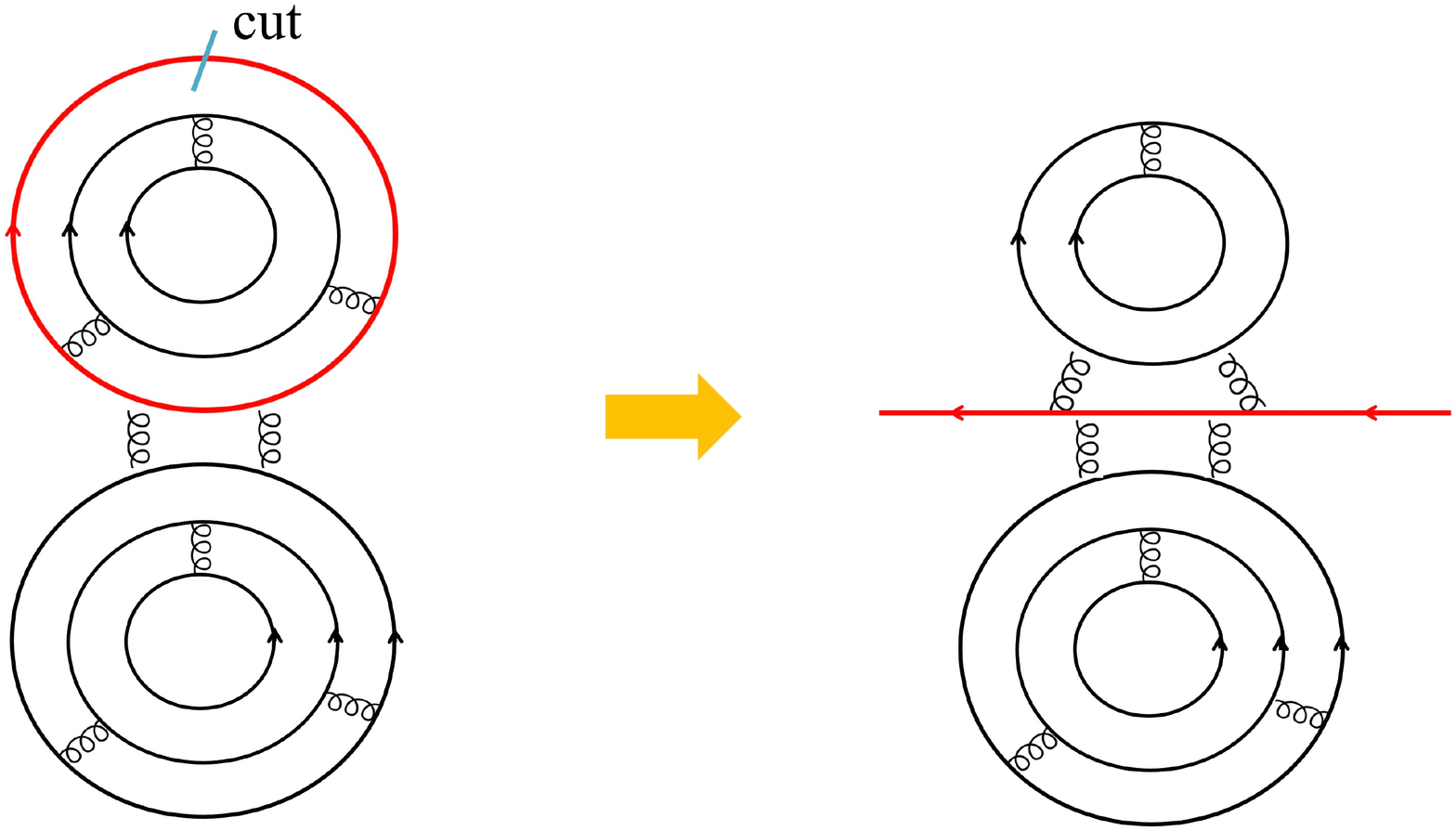}
\vspace{-1.3cm}
\caption{ \footnotesize{
A self-energy graph for a quark participating in  baryon-baryon interactions.
The gluon exchanges can happen in both the color antisymmetric and symmetric channels.
		} }
		\vspace{-0.1cm}
\label{fig:cut_BB_int}		
\end{center}
\end{figure}

So far our discussions on EoS are entirely based on quasiparticle pictures, based on either quarks or baryons;
the interactions for quarks have been taken into account only indirectly,
 by demanding that quark momenta are distributed to some range as they should localize by confining effects.
Now we perturb the calculations presented in the previous sections.
We discuss interactions within quark descriptions.

In principle, one can construct EoS from the information of single particle propagators. 
The pressure at given $\mu_B$ is given by
\beq
\calP(\mu_B) = \int_0^{\mu_B} \rmd \mu_B' \, n_B (\mu'_B) \,,
\eeq
where $n_B$ can be expressed as ($\gamma_0$: the zero component of the Dirac matrices)
\beq
n_B (\mu_B) = \frac{1}{\, \Nc \,} \Tr[ S_q \blueflag{ \gamma_0 } ] \,,
\label{eq:2PI}
\eeq
where $S_q$ is the quark propagator and the trace runs over all quantum numbers. 
This relation is exact for whatever interactions; 
for example, in functional frameworks such as the 2PI action \cite{Luttinger:1960ua,Baym:1962sx,Cornwall:1974vz}, 
Eq.(\ref{eq:2PI}) always follows from self-consistent treatments of the quark self-energy and interactions.
As a result, the effects of interactions can be included into the self-energy of the propagator.
If we need to include the baryon-baryon interactions, one should write the corresponding 2PI graphs and consider all possible cuts of quark propagators to generate the self-energy graph,
see Fig.\ref{fig:cut_in_B} for a quark in a baryon and Fig.\ref{fig:cut_BB_int} for a quark participating in baryon-baryon interactions.

In this paper, we simply assume a phenomenological parametrization for a single quark energy.
We consider the form for a single particle energy,
\beq
E_{q} (p; f_q) = \sqrt{p^2 + M_q^2 \,} + \calV [ f_q ] \,,
\eeq
where $\calV$ is the contribution from interactions that may depend on $f_q$.

\begin{figure}[htb]
\begin{center}	
\vspace{-0.1cm}
	\includegraphics[width=8.6cm]{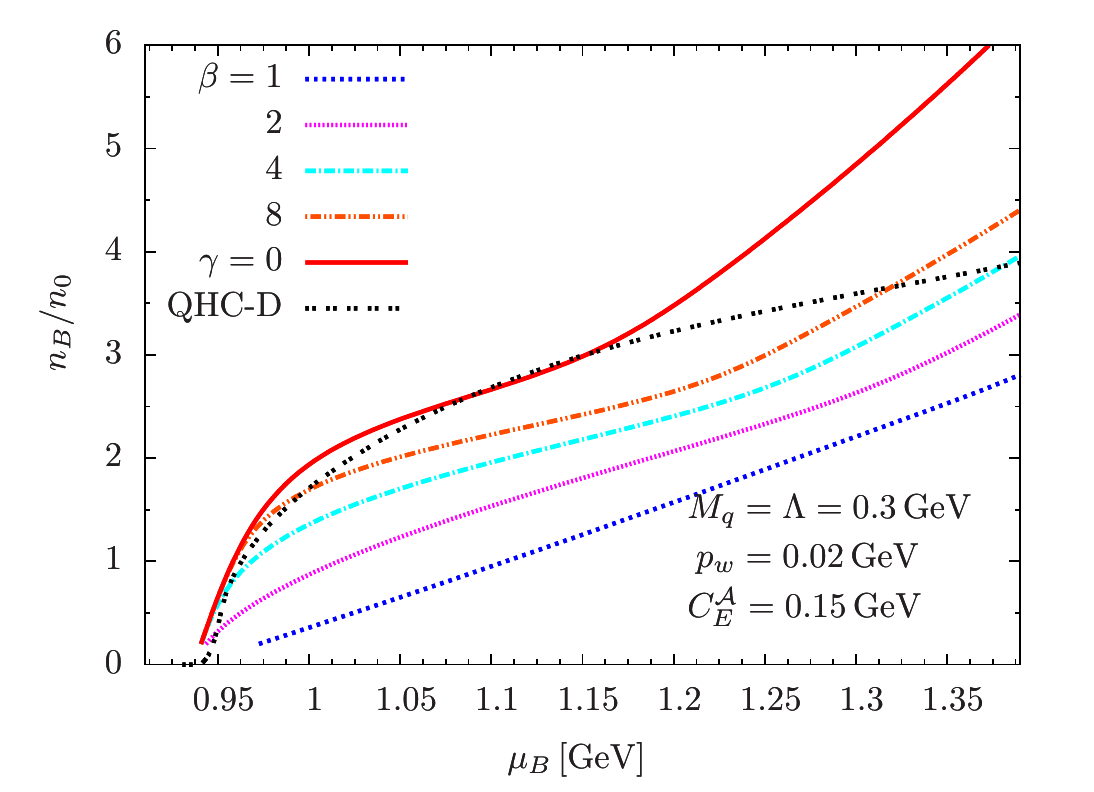}
	\includegraphics[width=8.6cm]{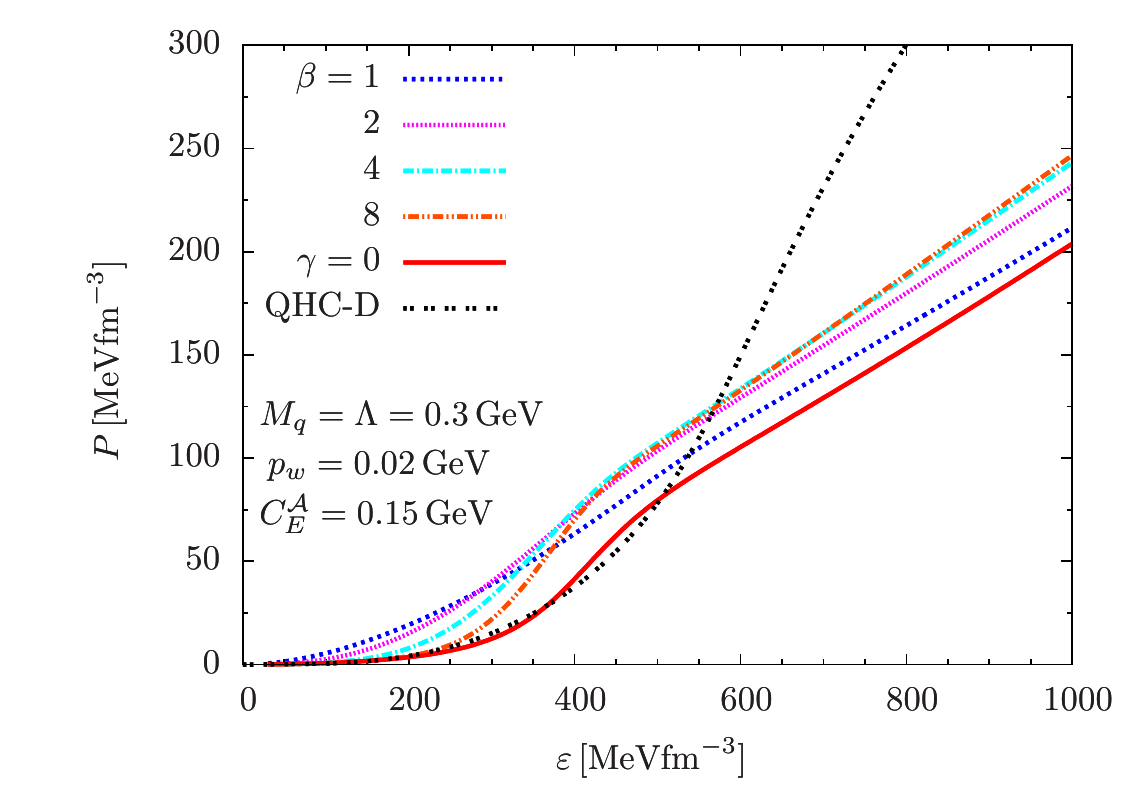}
	\includegraphics[width=8.6cm]{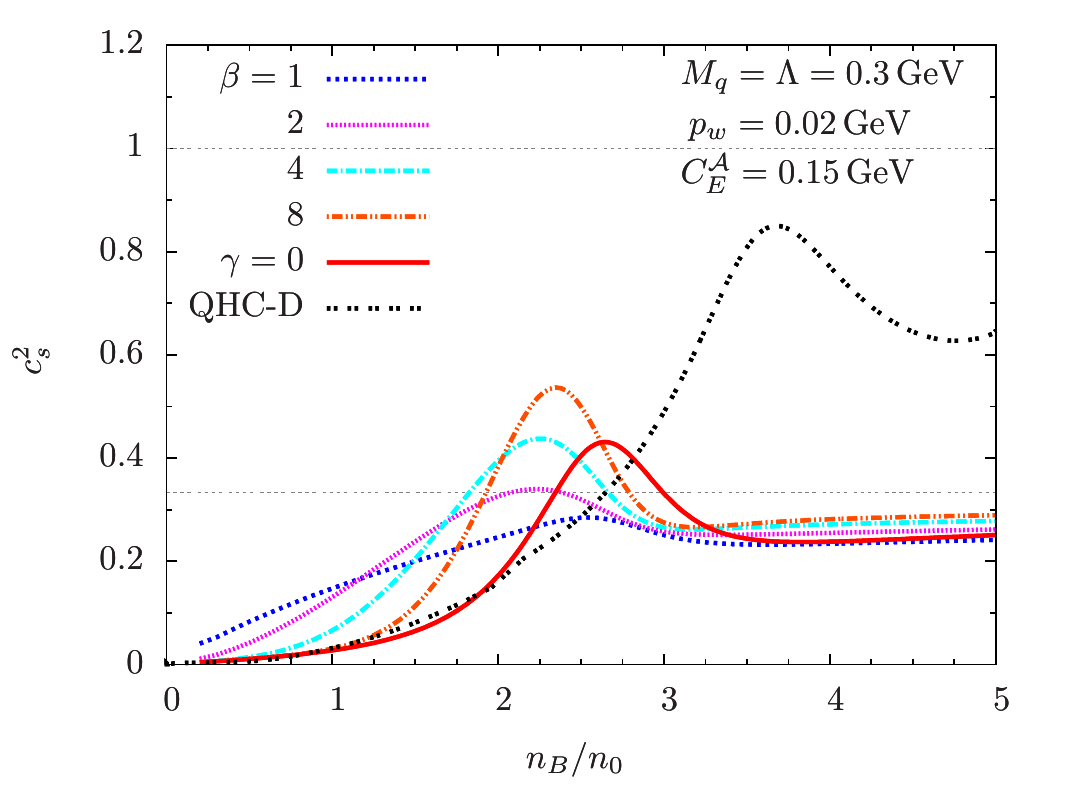}
\vspace{-0.4cm}
\caption{ \footnotesize{Equations of state, $n_B$ vs $\mu_B$, $\calP$ vs $\varepsilon$, and $c_s^2$ vs $n_B$, with $C_E^{\calA}=0.15$ GeV and $C_E^{\calS}=0$. 
The impacts of partially filled states are examined for the power $\beta=1,2,4,8$ with $\gamma=1$ or $\gamma=0$.
(Reminder: for the QHC19-D, $M_{\rm max}\simeq 2.28M_\odot$, $R_{1.4} \simeq 11.6$ km, and $R_{2.08}\simeq 11.5$ km.)
		} }
		\vspace{-0.8cm}
\label{fig:eos_int_vcs0}		
\end{center}
\end{figure}

\begin{figure}[htb]
\begin{center}	
\vspace{-0.1cm}
	\includegraphics[width=8.6cm]{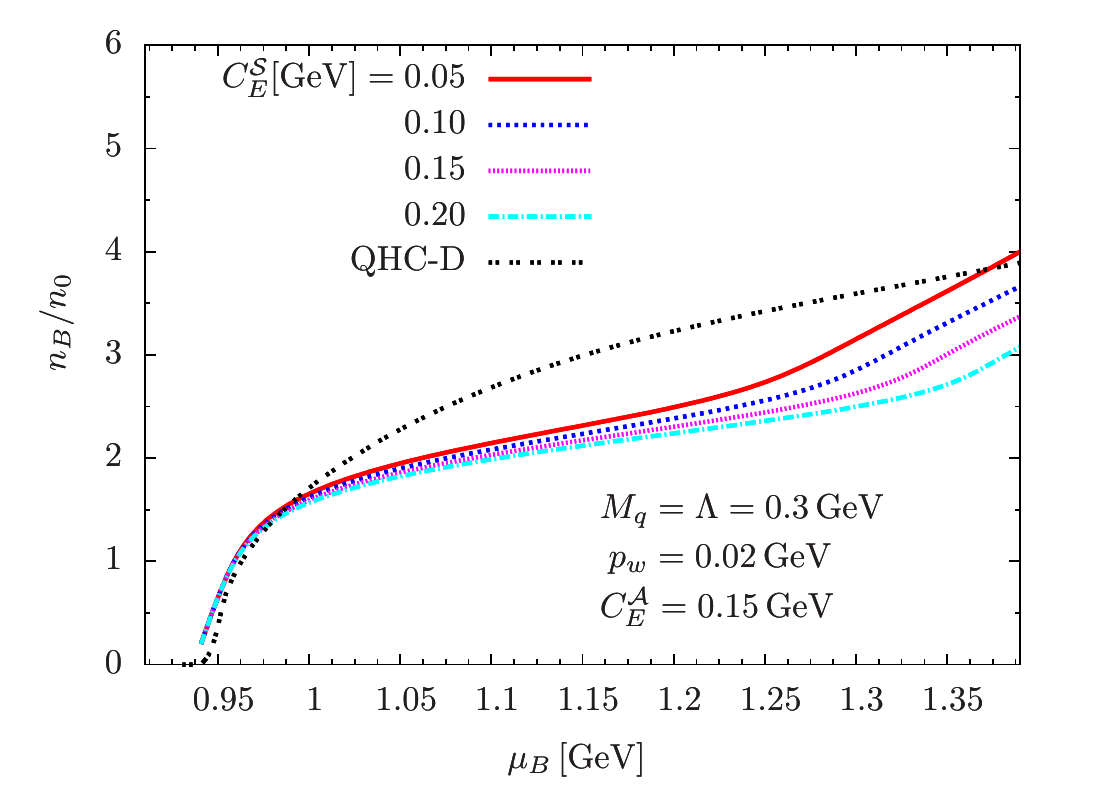}
	\includegraphics[width=8.6cm]{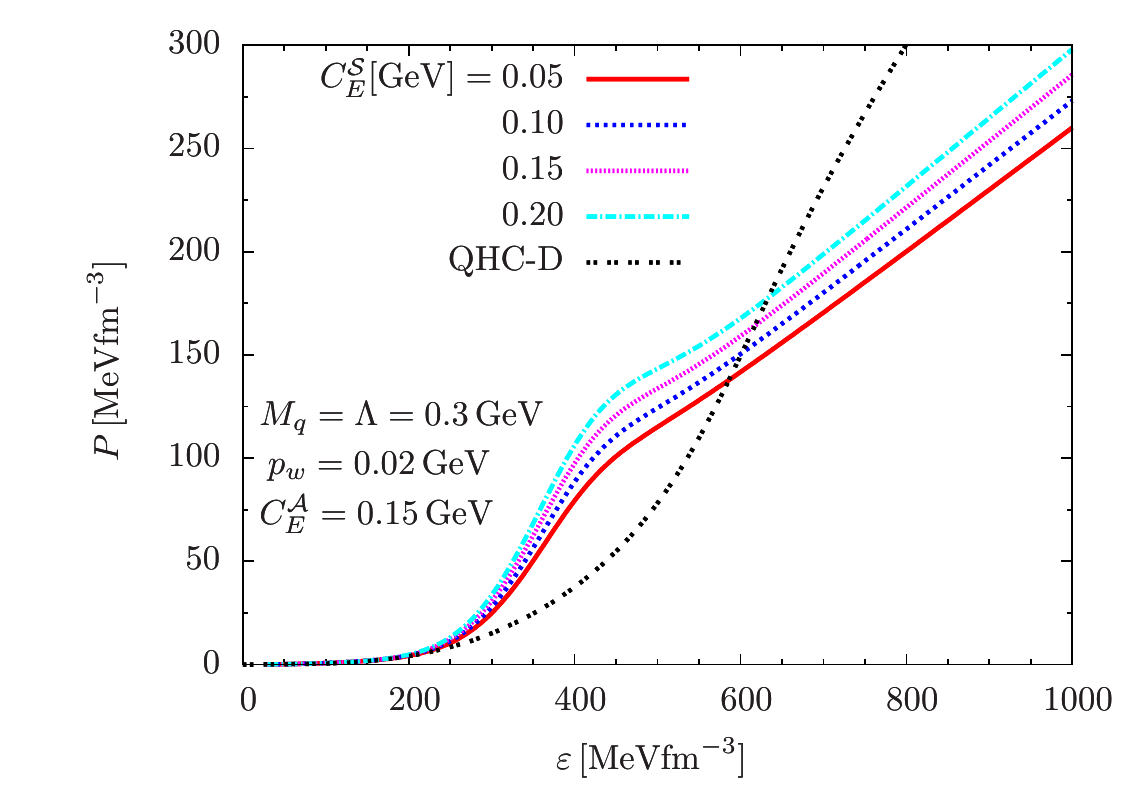}
	\includegraphics[width=8.6cm]{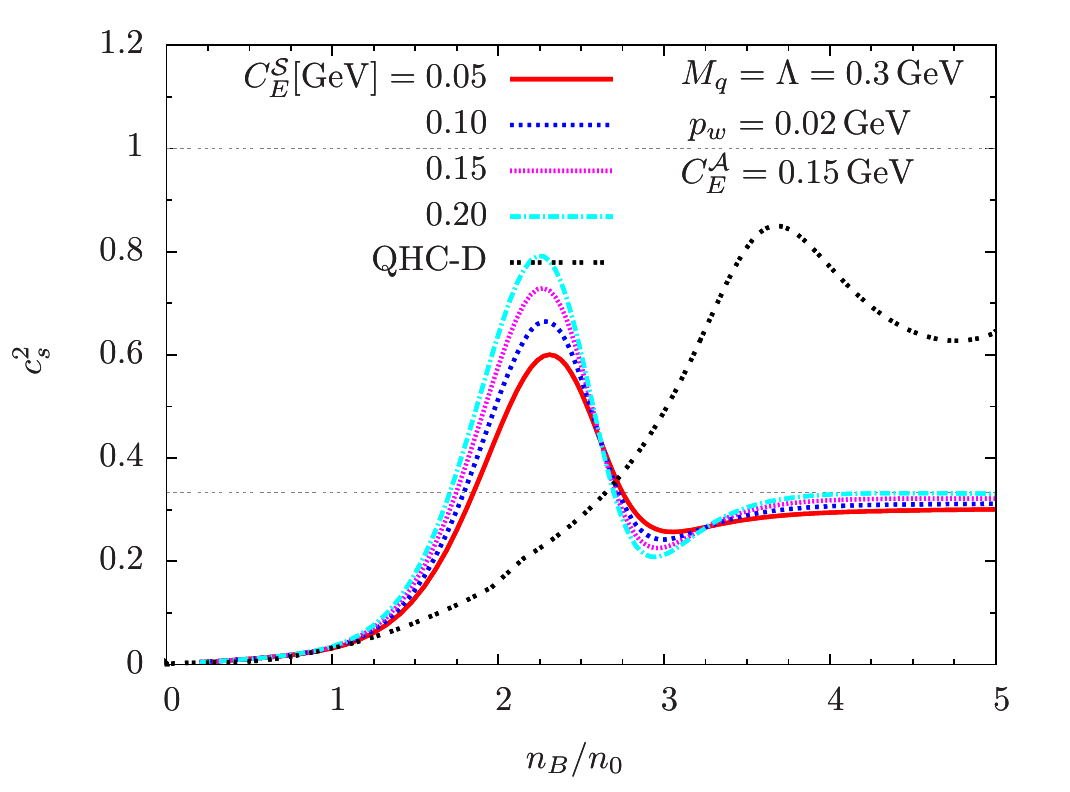}
\vspace{-0.4cm}
\caption{ \footnotesize{ The plots are same as Fig.\ref{fig:eos_int_vcs0} except that $\beta=8$ is fixed and  $C_E^{\calS}$ is varied from $0.05$ to $0.20$ GeV. 
		} }
		\vspace{-0.8cm}
\label{fig:eos_int}		
\end{center}
\end{figure}

Our first task is to adjust the baryon mass.
As in a quark model with a phenomenological confining potential \cite{DeRujula:1975qlm,Isgur:1979be}, our baryon mass is too massive due to the kinetic energies of localized quarks.
To reproduce the observed baryon spectra, we need color-electric interactions for the overall reduction in the masses for baryons, 
and color-magnetic interactions to get the correct mass splitting, e.g., the $N$-$\Delta$ and $\pi$-$\rho$ splittings.
In this work, we focus on the electric interaction but neglect the details of mass splittings. In vacuum, we consider
\beq
\calV^{\rm vac}_{\rm CE} [f_q] = - C^{\calA}_E ~(\le 0)  \,,
\eeq
where $\calA$ indicates an antisymmetric representation in colors.
We adjust $\calV^{\rm vac}_{\rm CE} $ to set the baryon mass to the nucleon mass, $M_N \simeq 939$ MeV.

In a baryon, the color-electric forces always reduce the average quark energy, as the color wave function is always antisymmetric for any combinations of quarks (e.g., Fig.\ref{fig:cut_in_B}). 
For symmetric wave functions, the color-electric forces yield repulsive forces, and such channels are inevitable when several baryons come close together (e.g., Fig.\ref{fig:cut_BB_int}).

To take into account these attractive forces in the dilute regime and the repulsive forces in denser regime,
we consider the parametrization ($C^{ \calS}_E \ge 0 $)
\beq
\calV_{\rm CE} [f_q] = - C^{\calA}_E \times \big(\, 1 - \gamma  f^\beta_q\, \big) + C^{\calS}_E  f^\beta_q \,,
\eeq
where $\calS$ indicates symmetric channels in colors.
The power $\beta$ controls the impacts of partially occupied levels,
and the parameter $\gamma$ is chosen to either 0 or 1 to examine the effects of partially filled states.

In this model, quarks with $p \gg p_{\rm sat}$ are free from the saturated levels and have the energy reduction  ($f_q \ll 1$),
\beq
\calV_{\rm CE} [f_q] \simeq - C^{\calA}_E \,,
\eeq
as quarks in a baryon.
This feature may be interpreted as the attractive correlations near the Fermi surface; 
since the quark states are not fully occupied, quarks can arrange their wave functions to enhance the portion of color-antisymmetric channels as in an isolated baryon.
Meanwhile, quarks with $p \ll p_{\rm sat}$ have less freedom for such arrangements, 
and feel the overall repulsion ($f_q \simeq 1$  for the saturated levels), 
\beq
\calV_{\rm CE} [f_q] \simeq - C^{\calA}_E \times \big(\, 1 - \gamma  \, \big) + C^{\calS}_E   \,,
\eeq
due to the appearance of the repulsive channels.
The repulsive energy is activated only when $p_{\rm sat}$ becomes substantial.
The power $\beta $ determines how sharply one can distinguish the filled ($f_q\simeq 1$) and partially filled ($f_q < 1$) states.

Now, we put these ingredients into our numerical analyses.
First we set $C_E^{\calS}=\gamma=0$.
Shown in Fig.\ref{fig:eos_int_vcs0} are various EoS, $n_B$ vs $\mu_B$, $\calP$ vs $\varepsilon$, and $c_s^2$ vs $n_B$, together with the results of QHC-D as a guideline for EoS consistent with the NS observations \cite{Baym:2019iky}.
We set $C_E^{\calA}=0.15$ GeV to get $M_B \simeq 0.94 $ GeV at $n_B=0$. 
The EoS at low density behaves as those in QHC19-D, but starts to deviate from QHC19 around $n_B \sim 3n_0$ due to softening of our model at high density.
The speed of sound has the maximum at $\sim 2.6n_0$, mainly determined by our model parameter $\Lambda$ (or baryon size).
We emphasize that stiffening takes place in our model solely by the saturation effects.

To get stiffer EoS at high density, we set $\gamma=1$ to activate the effects of partially filled states.
With $\beta=1$, the low density part is much stiffer than the baryonic part in the QHC19 (which is the Togashi EoS \cite{Togashi:2017mjp}), contradicting with nuclear EoS at $n_B \lesssim n_0$.
Here, quarks lost the attractive energy too rapidly, leading to rapid growth in $\mu_B$ as $n_B$ increases.
One can delay the loss of attractive energies by increasing $\beta$ and suppressing $f_q^\beta$ terms for $f_q <1$. 
We found that, for $\beta \gtrsim 8$, our EoS gets along with the nuclear EoS for $n_B\lesssim 1.5n_0$.
Meanwhile, the high density part is considerably softer than QHC19-D.

In order to make the high density part stiffer, the simple way is to turn on $C_E^{\calS}$.
We increase from $C_E^{\calS} = 0.05$ GeV to $0.20$ GeV, and the corresponding results are shown in Fig.\ref{fig:eos_int}.
The repulsive forces are activated only when $f_q \simeq 1$, or $p_{\rm sat}$ is large, so leaving the low density part as before, but stiffening the high density part where $p_{\rm sat}$ is substantial.
The location of the maximum in the speed of sound is not very sensitive to the choice of $C_E^{\calS}$, but the height becomes larger for a greater $C_E^{\calS}$.
We also note that the width of the peak is sensitive to our choice of $p_w$, as in Fig.\ref{fig:fq_for_pbf_lam_sat}.

\section{Summary}
\label{sec:summary}

\begin{figure*}[htb]
\begin{center}	
\vspace{-2.0cm}
	\includegraphics[width=12.cm]{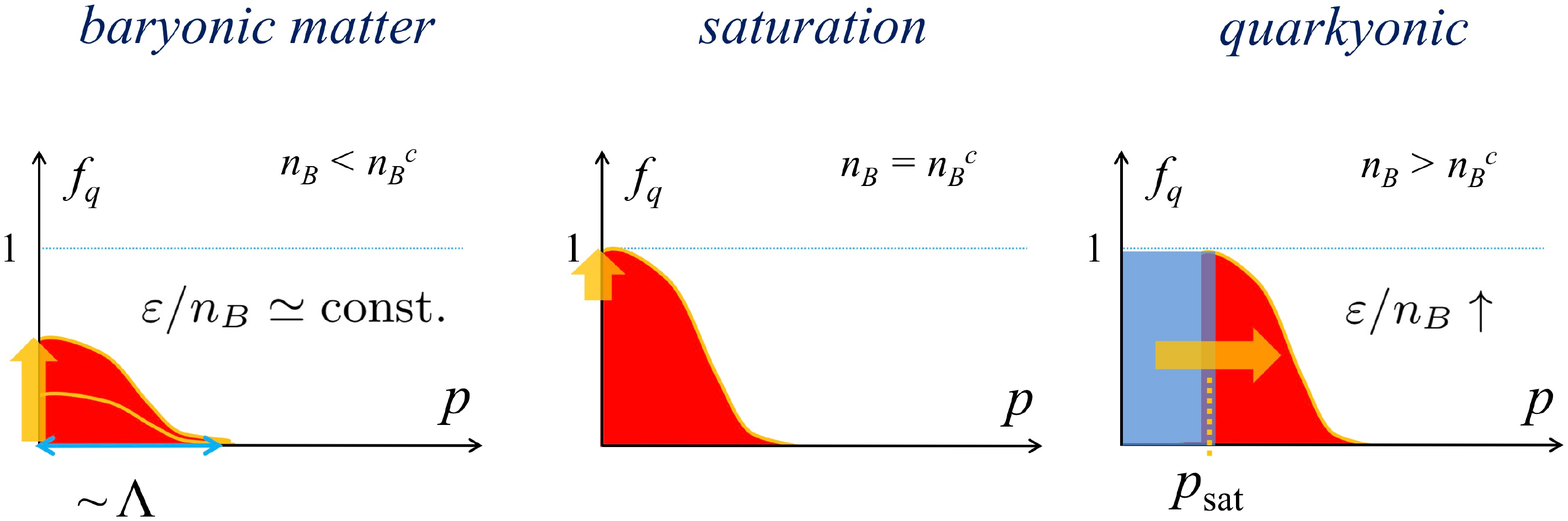}
\vspace{-2.2cm}
\caption{ \footnotesize{
A schematic picture for the evolution of matter from the baryonic to quark matter regime, see also Fig. 12 in Ref.\cite{Fukushima:2020cmk}.
In baryonic matter, the energy per particle is $\varepsilon/n_B \simeq $ const., and after the saturation, $\varepsilon/n_B$ starts to grow as in quark matter.
As the pressure is given by $\calP = n_B^2 \partial (\varepsilon/n_B)/\partial n_B$, matter after the saturation has much larger pressure than baryonic matter.
The speed of sound makes a peak around the density where the saturation happens.
		} }
		\vspace{-0.1cm}
\label{fig:quarkyonic_evolution}		
\end{center}
\end{figure*}

In this paper, we discussed how quark degrees of freedom stiffen EoS.
In order to relate the quark dynamics for a single baryon to baryonic matter and quark matter formation, 
we have introduced a model that relates three relevant functions: quark momentum distribution $Q_{\rm in}$ in a baryon, occupation probability of states for baryons $\calB$, and occupation probability of quark states $f_q$.
We also consider the effects of interactions at the quark level.

Below we summarize our findings:

(i) In a dilute regime, the confined quarks contribute to the energy density through the mass of baryons, but do not directly contribute to the pressure;
hence, the EoS are very soft (Fig.\ref{fig:quarkyonic_evolution} left).
This dilute regime continues until the low momentum states for quarks are saturated (Fig.\ref{fig:quarkyonic_evolution} middle).
The saturation can take place considerably before the baryons fully overlap, possibly at density close to the nuclear saturation density, $n_B \sim 1-3n_0$.
This picture is in line with the recent proposal of Soft Deconfinement as the onset of the mode-by-mode percolation \cite{Fukushima:2020cmk}.

(ii) After the saturation, the energy per particle, $\varepsilon/n_B$, begins to change as in quark matter (Fig.\ref{fig:quarkyonic_evolution} right), 
and the pressure, $\calP = n_B^2 \partial( \varepsilon/n_B )/\partial n_B$, grows rapidly,
although changes in $n_B$ and $\varepsilon$ are modest. 
These features lead to a peak in speed of sound, $c_s^2=\partial \calP/\partial \varepsilon$.
In our model, such a peak follows by just assuming the continuity of $f_q$ before and after the saturation, while no detailed descriptions of interactions were necessary.

(iii) To take into account the constraint $f_q \le 1$ after the saturation, it is easiest to directly work with the quark description.
But we can also infer the baryon momentum distribution $\calB$ consistent with the desired form of $f_q$.
Through such attempts, we reached a model similar to the MR model for quarkyonic matter. 
The resulting $\calB$ differs from the pure baryonic descriptions in which baryons are treated as if elementary particles;
baryons after the saturation are highly relativistic. 

(iv) We do not need many baryon species to fill the quark levels. 
For three flavors, we need $2\Nf$ baryon states to fill the quark color-flavor-spin states for a given spatial wave function.
In this respect, nucleons and hyperons should not be treated as independent when they share the same quark 
states \cite{Kojo:2015fua,Duarte:2020kvi}.

(v) While the stiffening of matter near the saturation seems a generic trend in our modeling,
a model without interactions does not lead to sufficient stiffness at high density, $n_B \gtrsim 4-5n_0$, 
that is required by the existence of $2M_\odot$ NSs.
This observation is consistent with viewpoints in our previous works \cite{Baym:2019iky,Song:2019qoh,Suenaga:2019jjv} where various short-range interactions of $p=0.2-1$ GeV were discussed.
One way to stiffen the high density part is to use a model in which 
quarks in saturated states feel repulsions but those near the Fermi surface enjoy the attractive correlations.

Unfortunately, our discussions remain largely qualitative and the treatments of dynamics are in many senses {\it ad hoc}.
There remain many things to be done, as listed below.

First, it is better to directly use a quark wave function in a constituent quark model.
To get $Q_{\rm in }$ in this work, we should first calculate the three-body wave function and then integrate out two momentum variables.
By doing this the matter properties can be directly expressed by quantities in hadron spectroscopy.

Second, to use the framework for predictions of the NS properties, we need to include the flavor asymmetry, and, in particular, have to discuss how nucleons fill the quark states at low energy.

Third, as we use a quark model, it is desired to directly use interactions in a quark model for baryons.
There have been many works to reproduce baryon properties, and the lattice QCD studies of baryon-baryon interactions support the idea that the short range part (such as the hard core repulsion among nucleons) is overall consistent with the descriptions based on quark dynamics with one-gluon exchanges.

Fourth, pairing effects leading to the chiral condensates or diquark condensates should be discussed to determine the phase structures as well as EoS.
In this work, we fixed the constituent quark mass as in vacuum, but it is very likely the effective mass changes with density.

Finally, the model should be extended to finite temperatures.
When we come to thermal excitations, we need to address whether thermal excitations are hadrons or quarks.
In fact, this is a crucial step to establish the quark-hadron continuity or the quarkyonic matter scenario, 
because the response in EoS as well as in the transports to changes in temperature is entirely different for hadronic and quark excitations \cite{Kojo:2020ztt}.
In two-color QCD, there is a hint from the lattice QCD for hadronic excitations at high density \cite{Kojo:2021knn,Suenaga:2021bjz}, although more data are needed to establish this idea.

In forthcoming papers we plan to give more quantitative estimates on EoS, arranging the setup for the baryon spectra and NS phenomenology.

\section*{Acknowledgments}

I thank  Larry D. McLerran and Robert D. Pisarski for having introduced me to the topic of quarkyonic matter;
Dyana C. Duarte, Saul Hernandez-Ortiz, Kie Sang Jeong for instructions about quarkyonic matter equations of state;
Kenji Fukushima and Wolfram Weise for discussions on Soft Deconfinement;
Gordon Baym and Tetsuo Hatsuda for general discussions on neutron star equations of state;
and Daiki Suenaga for discussions on thermal excitations in quark matter.
This work is supported by NSFC Grant No. 11875144.

\appendix

\section{Phase space density in baryonic bases}

In Sec.\ref{sec:occ_baryon},
 we considered the shell distribution of baryonic states after the quark saturation and have derived a constraint $h\Delta \lesssim \Lambda/\Nc^2$.
 The constraint acts on the product $h \Delta$, not on each of $h$ and $\Delta$ separately.
In the MR model, a thin shell structure with $h=1$ and $\Delta \sim \Lambda/\Nc^2$ is used.
Here we consider another possible choice.
This illustrates the peculiar structures of the baryon occupation probability after the quark saturation.

\begin{figure}[tb]
\begin{center}	
\vspace{-0.0cm}
	\includegraphics[width=8.cm]{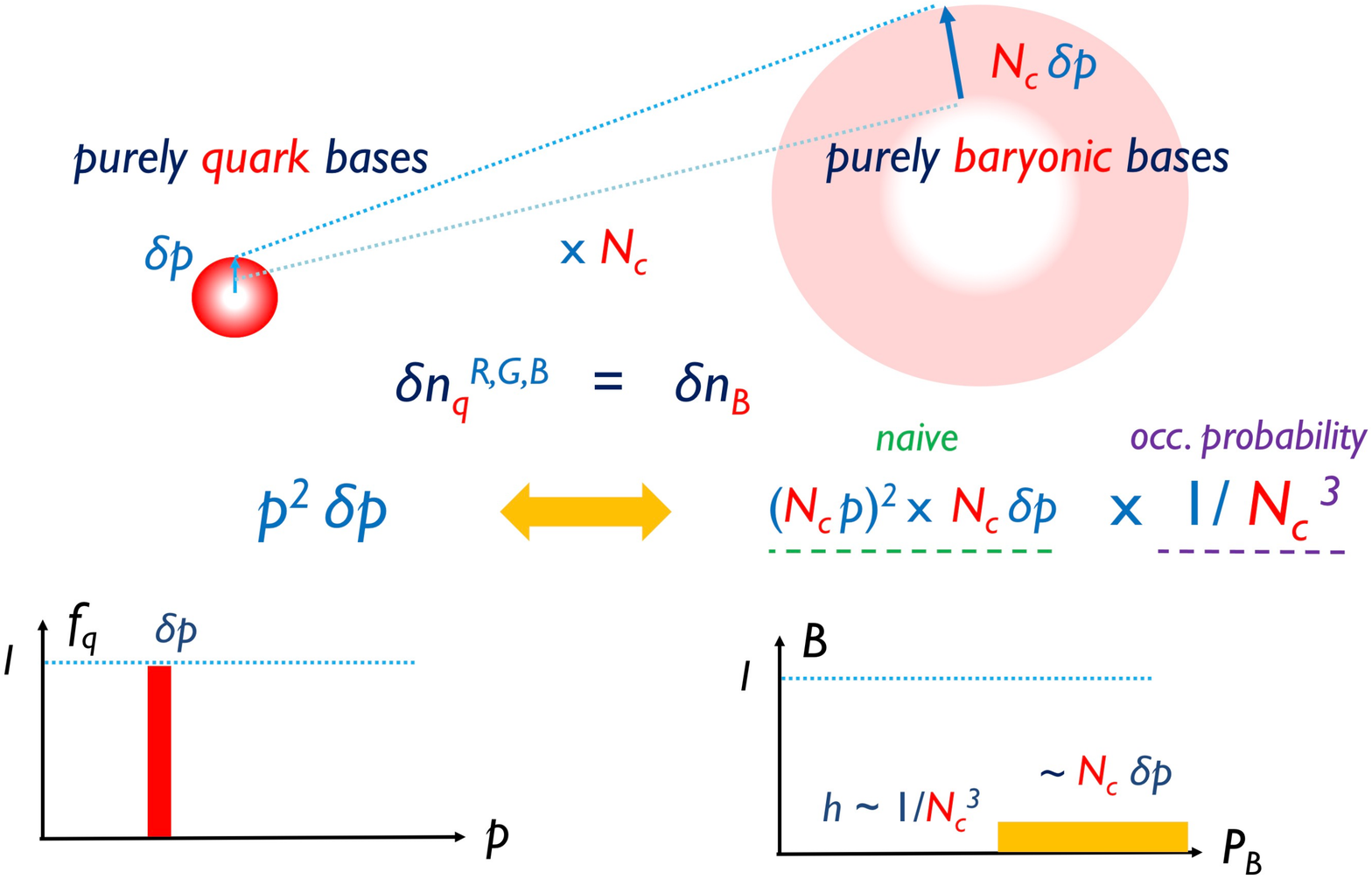}
\vspace{-0.cm}
\caption{ \footnotesize{
The comparison of the occupation probabilities for quark and baryon states.
We increase the quark  Fermi momentum $p$ by $\delta p$, with which the quark density for a given color (and spin-flavor) increase by $\delta n_q^{R,G,B} \sim p^2 \delta p$.
In terms of the purely baryonic bases, the baryon number must increase by the same amount, $\delta n_B = \delta n_q^{R,G,B}$.
In order to satisfy this consistency condition,
we have to conclude that the phase space in baryonic bases is more dilute than the quark's phase space. 
		} }
		\vspace{-0.1cm}
\label{fig:phase_space_density}		
\end{center}
\end{figure}

Let us discuss the increase of baryon number in two different bases, i.e., quark and baryon bases, see Fig.\ref{fig:phase_space_density}.
Noting the relation $n_B = n_q^{R,G,B} = n_q/\Nc$, 
we compare the change of quark Fermi sea for a specific color with that of the baryon Fermi sea.

In quark descriptions, we increase the quark Fermi momentum from $p$ to $p+\delta p$. The density increases as
\beq
\delta n_B = \delta n_q^{R,G,B} \sim p^2 \delta p \,,
\eeq
where $\Nc$ factors do not show up.

Now we try to describe the same increase in density using purely baryonic bases.
We note that the quark and baryon momenta near the Fermi surface are related as $P_B \sim \Nc p$.
Similarly $\delta P_B \sim \Nc \delta p$. 
Then, naively one would reach
\beq
\delta n_B^{\rm naive} \sim ( \Nc p )^2 \times \Nc \delta p = \Nc^3 p^2 \delta p\,, 
\eeq
which does not match with $\delta n_q^{R,G,B}$, due to the factor $\Nc^3$.
In order to avoid this contradiction, we have to conclude that the phase space for baryons is more dilute than the quark's phase space.
The proper estimate should take into account the occupation probability $h$ for baryonic states \cite{Kojo:2019raj},
\beq
\delta n_B \sim h \, \delta n_B^{\rm naive} ~\rightarrow~  h\sim 1/\Nc^3 \,.
\eeq
In terms of the $h\Delta$ constraint discussed before, 
here we are thinking of the case that $h \sim 1/\Nc^3$ and $\Delta \sim \Nc \delta p$.

Within this description, we consider the variation of energy density with respect to the change $\delta p$, 
and examine how the expression appears consistent for quark and baryon descriptions.
For simplicity, we assume the case where $p \gg \Lambda$ so that quarks and baryons have the energies $\sim p$ and $\sim \Nc p$, respectively. 

In quark bases, we need to sum contributions from all colored quarks,
\beq
\delta \varepsilon^{\rm quark} \sim \Nc \times p^3 \delta p \,.
\eeq
In baryon bases, we take into account the baryon occupation probability $h \sim 1/\Nc^3$ and get
\beq
\delta \varepsilon^{\rm baryon} \sim \frac{1}{\, \Nc^3 \,} \times (\Nc p)^3 ( \Nc \delta p) \,.
\eeq
So in both bases the variation of the energy density is $\sim \Nc p^3 \delta p$, as it should.
This represents the dual feature of quark and baryon descriptions;
the concept of occupation probability for baryonic states is the key to satisfy the consistency condition.

The remarkable feature in baryonic descriptions is that while baryons occupy the high energy states with small probability,
those states still make significant contributions to the baryon number and energy density.
This feature is difficult to be foreseen from pure baryonic considerations;
the quark saturation effects demand highly exotic descriptions for baryons.

\bibliographystyle{apsrev4-2}
\bibliography{ref}

\end{document}